\documentclass{sig-alternate}

\usepackage{graphicx}
\usepackage{balance}
\usepackage{amsmath}
\begin{document}
\newcommand{\QED}{\mbox{\rule[0pt]{1.0ex}{1.0ex}}}
\def\boxend{\hspace*{\fill} $\QED$}

\newtheorem{definition}{Definition}
\newtheorem{example}{Example}
\newtheorem{theorem}{Theorem}
\newtheorem{problem}{Problem}
\newtheorem{lemma}{Lemma}
\newtheorem{proposition}{Proposition}

\newcommand{\nop}[1]{}

\newcounter{line}
\input epsf

\hyphenation{op-tical net-works semi-conduc-tor}

\title{Performance Modeling and Evaluation for Information-Driven Networks}

\numberofauthors{2}

\author{
\alignauthor
Kui Wu\\
       \affaddr{Department of Computer Science}\\
       \affaddr{University of Victoria}\\
       \affaddr{B.C.  Canada V8W 3P6}\\
       \email{wkui@cs.uvic.ca}
% 2nd. author
\alignauthor
Yuming Jiang, Guoqiang Hu\\
      \affaddr{Q2S Center of Excellence}\\
       \affaddr{Norwegian University of Science and Technology}\\
       \affaddr{Trondheim, Norway}\\
       \email{\{jiang,guoqiang.hu\}@q2s.ntnu.no}
 }

%\date{}
\CopyrightYear{2008}
\maketitle

\begin{abstract}
Information-driven networks include a large category of networking systems, where network nodes are aware of information delivered and thus can not only forward data packets but may also perform information processing. In many situations, the quality of service (QoS) in information-driven networks is provisioned with the redundancy in information. Traditional performance models generally adopt evaluation measures suitable for packet-oriented service guarantee, such as packet delay, throughput, and packet loss rate. These performance measures, however, do not align well with the actual need of information-driven networks. New performance measures and models for information-driven networks, despite their importance, have been mainly blank, largely because information processing is clearly application dependent and cannot be easily captured within a generic framework. To fill the vacancy, we present a new performance evaluation framework particularly tailored for information-driven networks, based on the recent development of stochastic network calculus. We analyze the QoS with respect to information delivery and study the scheduling problem with the new performance metrics.  Our analytical framework can be used to calculate the network capacity in information delivery and in the meantime to help transmission scheduling for a large body of systems where QoS is stochastically guaranteed with the redundancy in information.  
\end{abstract}

% A category with the (minimum) three required fields
%A category including the fourth, optional field follows...
%\category{D.2.8}{Software Engineering}{Metrics}[complexity measures, performance measures]



\section{Introduction}\label{sec:introduction}
%What is Information-Driven Networks? What is the special chance we can take
Although computer networks in general are purposed for information delivery, most existing network architectures like the Internet are actually not information driven in the sense that network nodes (e.g., routers and switchers) only care about packets instead of the information inside. As a common principle, network nodes as well as the whole network system are designed to support quality of service (QoS) with respect to packet-oriented service measures such as bounded packet delay and promised data throughput. To achieve this, QoS provisioning mechanisms~\cite{Arm} have been proposed and used in the Internet, including admission control, congestion control, resource reservation, QoS routing, and so on. It has been observed that on the one hand QoS provisioning mechanisms provide certain service guarantee to privileged data traffic; on the other hand they largely increase the system complexity and incur a heavy burden on network nodes. Many emerging network systems, for example, wireless sensor networks, consist of nodes with only very limited computational capability and thus do not have the luxury  to accommodate complex QoS mechanisms. Nevertheless, QoS is important in any means. For instance, in a patient-monitoring system or a fire alarm system with wireless sensor networks, we certainly require important information like abnormal heart beats or high temperature readings to be delivered correctly to a monitoring center. The dilemma we face is to guarantee QoS maybe without any underlying promise from network nodes on timely per packet delivery.   

The traditional meaning of QoS, e.g., for guaranteed per packet delivery and end-to-end delay, is actually an overkill, since all we care is information. The traditional solutions focusing on packets instead of the information inside have the historical reason: the network protocol stack is layered and network protocols should not mix up with application-layer information. With the emergence of new technologies such as wireless sensor networks, however, the layering principle is not necessarily a rule of thumb, and the redundancy in the information sources should be utilized in network protocol design. A network node may not be purely a data forwarding device. Instead, it may become aware of information forwarded and is able to perform information processing whenever necessary. The ultimate goal of the whole network system is no longer to guarantee service for individual packets, but to guarantee a certain amount of information to be successfully transported. We call this type of networking systems \textit{information-driven networks}. Typical examples include wireless sensor networks with directed data diffusion~\cite{Int}, distributed content sharing over peer-to-peer networks~\cite{Ahm,Kun}, and networks using network coding~\cite{Ahl00,Yeung06}.  

\begin{figure}[tp]
 \centerline {\epsfxsize = 2.5 in \epsfbox{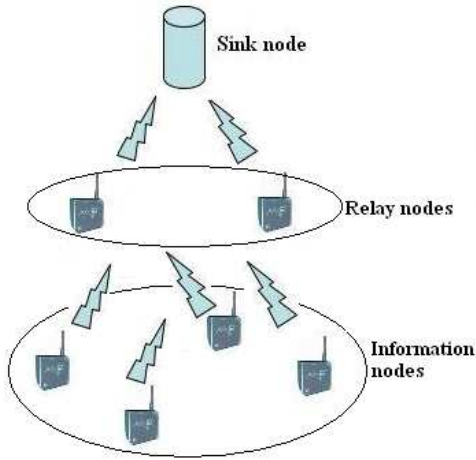}}
  \caption{A simple example of wireless sensor networks}\label{fig:fig1}
\end{figure}

Making the network to be information driven opens special opportunities for QoS provisioning, e.g., in environment where the network is subject to high packet losses or network nodes are stringently constrained by computational power and limited bandwidth. In applications where data exhibit spatial and/or temporal correlation, it is unnecessary to provide reliable transmission for each individual packet. Instead, QoS is guaranteed as long as required information can be obtained as sure (i.e., with a very high probability). We use a simple example to illustrate the advantages of allowing the network to become information aware. 
\begin{example}\label{ex:motivation}
Assume that a wireless sensor network includes six sensor nodes and one processing center, also called the sink node, as shown in Figure~\ref{fig:fig1}. Four sensors at the bottom of the figure monitor the environment and periodically send out measurement data like temperature, humidity readings. Two sensors in the middle of the figure are used as data relay to the sink. Wireless links are generally subject to a high loss rate in wireless sensor networks~\cite{Sze}, so we assume that the average packet loss rate is $25\%$ for each wireless link. Without considering information, we treat the network as purely a data delivery system like the Internet. In this case, we need to make sure that each data packet is correctly delivered from the source to the sink with a high probability. If we set this probability to be no smaller than $96\%$, we need about $24$ transmissions in total (calculated with two retransmissions each link to guarantee the high probability of correct end-to-end packet delivery). In contrast, if relay nodes know that the information from the four source sensors is highly correlated and if the information is considered to be delivered as long as at least one packet from the sources is received by the sink, eight transmissions (e.g., without any retransmissions) can guarantee that the information is delivered with a probability no smaller than $96\%$. 
\end{example} 

The above example clearly illustrates the necessity of taking information into consideration. Yet, several difficulties need to address even in the very simple example. First, how can we capture and model the correlation at the information sources? In addition, the correlation may change over time. How can we capture the dynamic changes in a timely fashion? Second, the above example only considers the correlation at the information sources, how can we perform information processing at intermediate relay nodes for better QoS provisioning and resource saving? Third, the use of application-layer information in network protocols has changed the fundamental design principle of current Internet architecture, where the network is considered as a packet transportation tool and the service guarantee is promised for individual packets. This fundamental change renders traditional performance modeling and evaluation approaches invalid for information-driven networks. For instance, network throughput in terms of number of bits per time unit and end-to-end packet delay are no longer good measures and new metrics should be used to align with the need of information-driven networks. What should be a good model for performance evaluation and resource scheduling for information-driven networks?

During the last several years, there are substantial research efforts devoted to tackling the first two difficulties. Particularly, the spatial and temporal correlations of information have been studied and utilized in network scheduling and resource saving in wireless sensor networks~\cite{Liu,Pat,Vur}; information redundancy has been exploited to help load balancing and improve fault tolerance in peer-to-peer content sharing systems~\cite{Ahm,Kun}. Regarding the third challenge, to the best of our knowledge, the only attempts to accommodate information processing in performance modeling are the work in~\cite{Fidler06,Sch}. Nevertheless, information processing is simply modeled with a scaling function in~\cite{Fidler06,Sch} and the information embedded in data packets has not been modeled, let alone utilized. In this sense, the performance models in~\cite{Fidler06,Sch} are not really information-driven models. A systematic performance modeling framework suitable for information-driven networks still remains largely open. 

%Our idea and main contributions
 In this paper, we propose the first-of-the-kind analytical framework suitable for performance study and resource scheduling of information-driven networks. Particularly, we make the following contributions:
 \begin{enumerate}
\item We present a comprehensive performance model for information-driven networks based on the recent development of stochastic network calculus. Particularly, our model captures the information correlation and the QoS guarantee with respect to stochastic information delivery rates, which have never been formally modeled before. 
\item We study the stochastically achievable network capacity in information delivery and derive the bounds on probability of information delivery within a given end-to-end information delay. 
\item We study the transmission scheduling problem in information - driven networks and design algorithms to search for feasible transmission schedules to meet given QoS requirements. 
\item We give examples to illustrate how our analytical framework can be used to solve problems in practice.  
\end{enumerate}

The rest of the paper is organized as follows. Section~\ref{sec:background} includes the background knowledge on stochastic network calculus. In Sections~\ref{sec:informationCalculus} and~\ref{sec:model}, we present a calculus and a system model for information-driven networks, respectively.  We analyze the stochastically achievable information delivery rates in Sections~\ref{sec:analysis}, and study the transmission scheduling problem in Section~\ref{sec:scheduling}. We present the numerical experiment results in Section~\ref{sec:evaluation}. In Section~\ref{sec:furtherDiscussion} we answer several common questions on information-driven network calculus, and in Section~\ref{sec:work} we introduce related work. The paper is concluded in Section~\ref{sec:conclusion}.  

%% Background Knowledge
\section{Background of Stochastic Network Calculus} \label{sec:background}
\subsection{Notation}
We first introduce the notation and key concepts of stochastic network calculus~\cite{Jia,Li}. Throughout this paper, we assume that all arrival curves and service curves are non-negative and wide-sense increasing functions. Following the convention, we use $A(t)$ and $A^*(t)$ to denote the cumulative traffic arrives and departures in time interval $(0,t]$, respectively, and use $S(t)$ to denote the cumulative amount of service provided by the system in time interval $(0,t]$. For any $0\le s \le t$, we denote $A(s,t) \equiv A(t)-A(s), A^*(s,t) \equiv A^*(t)-A^*(s),$ and $S(s,t) \equiv S(t)-S(s).$ We adopt $A(0)=A^*(0)=S(0)=0$. 

We denote by $\mathcal{F}$ the set of non-negative wide-sense increasing functions, i.e.,  $$\mathcal{F}= \{f(\cdot): \forall 0\le x\le y, 0\le f(x) \le f(y)\},$$ and by $\bar{\mathcal{F}}$ the set of non-negative wide-sense decreasing functions, i.e.,  $$\bar{\mathcal{F}}= \{f(\cdot): \forall 0\le x\le y, 0\le f(y) \le f(x)\}.$$ 

For any random variable $X$, its distribution function, denoted by $F_X \equiv Prob\{X\le x\}$, belongs to $\mathcal{F}$, and its complementary distribution function, denoted by $\bar{F}_X \equiv Prob\{X>x\}$, belongs to $\bar{\mathcal{F}}$. 

\subsection{Operators}
The following operations defined under the $(\min,+)$ algebra~\cite{Chang00, Cruz91a,Le} will be used in this paper: 
\begin{itemize}
\item The $(\min,+)$ \textit{convolution} of functions $f$ and $g$ is 
\begin{equation}
(f\otimes g) (t) = \inf_{0\le s \le t}\{f(s) + g(t-s)\}.
\end{equation}
 \item The $(\min,+)$ \textit{deconvolution} of functions $f$ and $g$ is 
\begin{equation}
(f\oslash g) (t) = \sup_{s\ge0}\{f(t+s) - g(s)\}.
\end{equation}
\item The $(\min,+)$ \textit{inf-sum} of functions $f$ and $g$ is:
\begin{equation}
(f\odot g) (t) = \inf_{s\ge0}\{f(t+s) + g(s)\}.
\end{equation}
\end{itemize}
\nop{Note that the $\odot$ operator has not been used before. It is defined in this paper to simplify notation. In addition, we need the normal convolution in our analysis:
\begin{itemize}
\item The \textit{normal convolution} of functions $f$ and $g$ is 
\begin{equation}
(f * g) (x) = \int_0^x f(x-y)dg(y).
\end{equation}
\end{itemize}}
%In addition, we adopt:
%\begin{itemize}
%\item $[x]^+ \equiv max\{x,0\}$.
%\end{itemize}

\subsection{Performance Measures, Traffic and Server Models}
%\subsection{Performance Measures and Stochastic Curves}
The following measures are of interest in service guarantee analysis under network calculus:
\begin{itemize}
\item The backlog $\mathcal{B}(t)$ in the system at time $t$ is defined as:
\begin{equation}
\mathcal{B}(t) = A(t) - A^*(t).
\end{equation}
\item The delay $\mathcal{D}(t)$ at time $t$ is defined as:
\begin{equation}
\mathcal{D}(t)= \inf \{\tau \ge 0: A(t) \le A^*(t+\tau)\}.
\end{equation}
\end{itemize} 

Stochastic arrival curve and stochastic service curve are core concepts in stochastic network calculus with the former for traffic modeling and the latter for server modeling. It is worth noting that the deterministic arrival curve traffic model and the deterministic service curve server model under (deterministic) network calculus are a special case of their corresponding stochastic definition. In the literature, there are several definition variations of stochastic arrival curve and stochastic service curve \cite{Jia} such as:
\begin{definition}\label{def-sac}
A flow $A(t)$ is said to have a \textit{maximum-virtual-backlog-centric stochastic arrival curve} $\alpha \in \mathcal{F}$ with bounding function $f\in \bar{\mathcal{F}}$, denoted by $A\sim_{m.b.c.}<f,\alpha>$, iff for all $t \ge 0$ and all $x\ge 0$, there holds~\cite{Jia} 
\begin{equation}
Prob\{\sup_{0\le s \le t} \sup_{0\le u \le s} [A(u,s)-\alpha(s-u)] > x\} \le f(x).
\end{equation}
\end{definition}

\begin{definition}\label{def-ssc}
A server is said to provide a flow $A(t)$ with a \textit{stochastic service curve} $\beta \in \mathcal{F}$ with bounding function $g \in \bar{\mathcal{F}}$, denoted by $S\sim_{s.c}<g,\beta>$, iff for all $t\ge 0$ and all $x\ge 0$, there holds~\cite{Jia}  
\begin{equation}
Prob\{\sup_{0\le s \le t}[A\otimes\beta(s)-A^*(s)] >x\} \le g(x). 
\end{equation}
\end{definition} 

With the above definitions and their variations, various properties of stochastic network calculus, including stochastic backlog and stochastic delay bounds, have been proved (e.g., see~\cite{Jia,Li}).

\nop{ 
Among the many properties in stochastic network calculus~\cite{Jia2}, the following result is most useful in this paper~\cite{Jia}:
\begin{theorem} 
If the input flow $A\sim_{v.b.}<f,\alpha>$ and the system provides the input flow with service $S\sim_{s.c}<g,\beta>$, then 
\begin{enumerate}
\item The backlog $B(t)$ of the flow in the system at time $t$ satisfies: for all $t\ge 0$ and all $x\ge 0$, 
\begin{equation}
Prob\{B(t)>x\} \le f\otimes g(x+\inf_{x\ge 0}[\beta(s) - \alpha(s)]);
\end{equation}
\item The delay $D(t)$ of the flow in the system at time $t$ satisfies: for all $t\ge 0$ and all $x\ge 0$, 
\begin{equation}
Prob\{D(t)>h(\alpha+x, \beta^*)\} \le f\otimes g(x),
\end{equation}
where $\beta^*(t) = [\beta(t)-x]^+ + x$ and $h(\alpha,\beta)$ is the maximum horizontal distance between functions $\alpha$ and $\beta$ and is defined as 
$$h(\alpha,\beta)= \sup_{s\ge0}\{\inf\{\tau \ge 0: \alpha(s)\le \beta(s+\tau)\}\}.$$
\end{enumerate}
\end{theorem}
 }

\section{An Information-Driven Network Calculus}\label{sec:informationCalculus}
\subsection{Notation on Information} 
\label{sec:informationNotation}
Traditional stochastic network calculus uses the traffic arrival curve and the service curve to count for cumulative amount of traffic or service. If network nodes are information driven and can perform in-network processing, we need a translation from the amount of traffic to the amount of information in order to model information processing at network nodes. We hence develop a stochastic network calculus dedicated to performance modeling of information-driven networks.  

Although entropy (or entropy rate) is broadly used to measure information of random variable (or random process)~\cite{Cov}, to avoid various details in entropy estimation in specific applications~\cite{Lal}, we simply use the notation $H(A(t))$ to denote the information of $A(t)$ but leave \textit{its practical meaning and calculation open to users}. \nop{Clearly, the interested measure is application specific. For example, if we are interested in the temperature variation within a greenhouse, the interested measure in the flow may be temperature readings; if we are interested in the content distribution from different providers, the interested measure in the flow may be the IP addresses of contend providers.} Nevertheless, we need to define basic properties of information to make further analysis possible. 

\begin{definition} \label{def:information}The information of a flow $A(t)$ is denoted as $H(A(t))$ and has the following properties: 
\begin{enumerate}
\item $H(\emptyset)=0$ where $\emptyset$ denotes the null set; $H(A(0))=0$.
\item $H(A(t))$ is a non-negative, non-decreasing function of time $t$, and for $\forall t$ and $0\le s \le t$, $H(A(t)) = H(A(s))+ H(A(s,t))$ holds for the \textit{same} flow, i.e., the flow from the same information source.
\item For different flows $A_i(t), i=1,\ldots, N$, 
\begin{equation}
\sum_{i=1}^NH(A_i(t)) \ge H(\sum_{i=1}^NA_i(t)),
\end{equation}
where $\sum_{i=1}^NA_i(t)$ means the superposition of flows\\ $A_1(t), \ldots, A_N(t)$. 
\end{enumerate}
\end{definition}

\nop{
, is calculated with the entropy of an interested measure included in the flow, i.e., 
\begin{equation}
H(A(t)) = - \sum_{i=1}^{N(t)} P_i(t) \log P_i(t), \label{def:1}
\end{equation}  
where $N(t)$ is the total number of possible values of the interested measure, $P_i(t)$ is the probability that the $i$-th value occurs in $A(t)$.  
\end{definition}
}

The first property of information means that the information of a flow is initially zero. The second one means that the information of a flow does not become smaller as more data arrive and the information from the same flow is accumulative, i.e., $H(A(s,t))$ should be understood as the new information in the flow at time interval $(s, t]$. The third property means that the superposition of different flows do not increase the total information in the flows. Note that these three properties are consistent with the entropy definition. %The last property will be illustrated in more detail after we define the \textit{redundant information}. \nop{These three properties are consistent with the traditional entropy definition.}   

For ease of exposition, we omit the time index whenever doing so will not cause confusion.

\begin{definition} \label{def:redundant}
\textit{Redundant information} is a measure of the information redundancy in different flows, $A_1, \ldots, A_N$. It is defined in this paper as 
\begin{equation}
I(A_1;\ldots;A_N) \equiv \sum_{i=1}^NH(A_i) - H(\sum_{i=1}^NA_i), \label{eq:mututalInformation}
\end{equation}
and has the following properties: \nop{where $H(A_i)$ denotes the information of $A_i$ and $H(A_i|A_j)$ denotes the conditional information of $A_i$ if $A_j$ is known. The conditional information $H(A_i|A-j)$ is a $H$ function with the following extra properties:} 
\begin{enumerate}
\item If $A_1, \ldots, A_N$ are independent, $I(A_1;\ldots;A_N)$ equals $0$.
\nop{\item If $A_i$ and $A_j$ are identical, $I(A_i;A_j)$ reaches its maximum value, $H(A_i) (=H(A_j))$, which is the information of $A_i$ (or $A_j$).}
\item For any $i (=1, \ldots, N)$, $I(A_1;\ldots;A_N) \le H(A_i).$     
\end{enumerate}  
\end{definition}

%Based on Definition~\ref{def:redundant} , the third property of information in Definition~\ref{def:information} clearly holds.

\begin{definition} 
\textit{Information of a set of data sources} $A=<A_1, \ldots, A_M>$, $H(A)$, is defined as 
\begin{equation}
H(A) \equiv H(\sum_{i=1}^MA_i) = \sum_{i=1}^{M} H(A_i) - I(A_1;\ldots;A_M)  
\end{equation}
\end{definition}
\nop{
Again, the conditional information is just a notation with the above features, and its calculation is left to users. It is clear that when $A_i$ and $A_j$ are independent, $I(A_i;A_j)$ reaches its minimum value, $0$. When $A_i$ and $A_j$ are identical, $I(A_i;A_j)$ reaches its maximum value, $H(A_i)$, which is the information of $A_i$. }

\subsection{Modeling Flow and Service with Respect to Information}
\begin{definition}
A flow $A(t)$ is said to have an \textit{information stochastic arrival} curve $\alpha \in \mathcal{F}$ with bounding function $f\in \bar{\mathcal{F}}$, denoted by $A\sim_{i.s.a.}<f,\alpha>$, iff for all $t \ge 0$ and all $x\ge 0$, there holds 
\begin{equation}
Prob\{\sup_{0\le s \le t}\sup_{0\le u \le s} [H(A(u,s))-\alpha(s-u)] > x\} \le f(x).
\end{equation}
\end{definition}

\begin{definition}
A flow $A(t)$ is said to have an lower - bounded information stochastic arrival curve $\gamma \in \mathcal{F}$ with bounding function $\theta \in \mathcal{F}$, denoted by $A\sim_{l.i.s.a}<\theta,\gamma>$, iff for all $t \ge 0$ and all $x\ge 0$, there holds 
\begin{equation}
Prob\{\inf_{0\le s \le t}\inf_{0\le u \le s} [H(A(u,s))-\gamma(s-u)] \le x\} \le \theta(x).
\end{equation}
\end{definition}

\begin{definition}\label{def:service}
A server is said to provide a flow $A(t)$ with an \textit{information stochastic service} curve $\beta \in \mathcal{F}$ with bounding function $g \in \bar{\mathcal{F}}$, denoted by $S\sim_{i.s.s.}<g,\beta>$, iff for all $t\ge 0$ and all $x\ge 0$, there holds 
\begin{equation}
Prob\{\sup_{0\le s \le t}[H(A(s))\otimes[\beta]^{x}(s)-H(A^*(s))] >x\} \le g(x) 
\end{equation}
where $[\beta]^{x}(t) \equiv \max\{\beta(t), x\}$.
\end{definition}

Note that information service includes \textbf{both} information processing and information transmission. We also use the terms \textit{information service rate} and \textit{information arrival rate}: 
\begin{definition}\label{def:servicerate}
The (average) information service rate of a server following $\sim_{i.s.s.}<g,\beta>$ is defined as $\lim_{t \rightarrow \infty} \frac{\beta(t)}{t}$.
\end{definition}
\begin{definition}\label{def:arriveRate}
The (average) information arrival rate of a flow following $\sim_{i.s.a.}<f,\alpha>$ is defined as $\lim_{t \rightarrow \infty} \frac{\alpha(t)}{t}$.
\end{definition}

In the rest of the paper, unless otherwise mentioned, service by default is referred as information service.  

\subsection{Performance Measures}
The following definitions are used for information guarantee analysis:
\begin{definition}
The \textit{information delay} of input flow $A(t)$ in a system at time $t$ is defined as:
\begin{equation}
D(t) = \inf \{\tau \ge 0: H(A(t)) \le H(A^*(t+\tau))\},
\end{equation}
where $A^*(t)$ is the output flow. 
\end{definition}
\begin{definition}
The \textit{information backlog} at time $t$ in a system is defined as:
\begin{equation}
B(t) = H(A(t)) -H(A^*(t)),
\end{equation}
where $A(t)$ and $A^*(t)$ are input flow and output flow, respectively.
\end{definition}
\begin{definition}
The \textit{information backlog within delay bound} $\tau (\le D(t))$ at time $t$ is defined as:
\begin{equation}
\hat{B}(t,\tau) = H(A(t)) -H(A^*(t+\tau)).
\end{equation}
\end{definition}

\begin{figure}[tp]
 \centerline {\epsfxsize = 2.5 in \epsfbox{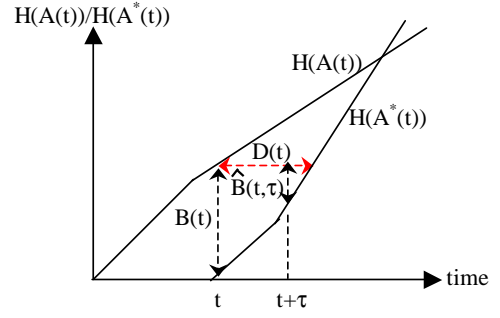}}
  \caption{The illustration of performance measures}\label{fig:illustration}
\end{figure}

To help better understanding, Figure~\ref{fig:illustration} illustrates the above three definitions. 

\subsection{Basic Properties of Information-Driven Network Calculus} 

To make the paper concise and easy to follow, we put all proofs in Appendix~\ref{app:A}.
 \nop{ \subsection{Lemmas}
The following lemmas will be used in this paper. The proof of Lemma~\ref{lemma:2} is provided in Appendix~\ref{app:A}.}

\begin{lemma}\label{lemma:1}~\cite{Jia} For any random variables $X$ and $Y$, and $\forall x \ge 0$, if $\bar{F}_X(x) \le f(x)$ and $\bar{F}_Y(x) \le g(x)$, where $f, g \in \bar{\mathcal{F}}$, then 
\begin{equation}
Prob\{X+Y >x\} \le (f\otimes g)(x). 
\end{equation}
\end{lemma}
\begin{lemma}\label{lemma:2} For any random variables $X$ and $Y$, and $\forall x \ge 0$, if $\bar{F}_X(x) \le f(x)$ and $F_Y(x) \le g(x)$, where $f\in \bar{\mathcal{F}}, g\in \mathcal{F}$, then 
\begin{equation}
Prob\{X-Y \ge x\} \le (f\odot g)(x). 
\end{equation}
\end{lemma}

\nop{\begin{lemma}\label{lemma:clipper}\textbf{(Output of information clipper)} Assume that the input flow and the output flow of an information clipper $C(t)$ are $A(t)$ and $A'(t)$, respectively. If $A\sim_{i.s.a.}<f,\alpha>$, then $A'\sim_{i.s.a.}<f,\alpha-H(C)>$. 
\end{lemma}
\textit{Proof.} The lemma is true because $H(A'(s,t)) - (\alpha(t-s)-H(C(s,t))) = H(A(s,t))-\alpha(t-s)$ \QED.}

\nop{
\begin{lemma}\label{lemma:output} \textbf{(Output of i.s.s.)} Consider a system with input flow $A$. If $A \sim_{i.s.a.}<f, \alpha>$, and the system provides the flow with service $S\sim_{i.s.s.}<g,\beta>$, then the output $A^*\sim_{i.s.a}<f\otimes g, \alpha\oslash\beta>.$
\end{lemma}
\textit{Proof.} Based on the properties of information, we have
\begin{align*}
&H(A^*(s,t)) = H(A^*(t))-H(A^*(s)) \\
\le & H(A(t)) - H(A^*(s)) \\
= &H(A(t)) - H(A(s))\otimes \beta(s) + H(A(s))\otimes \beta(s) \\
  &- H(A^*(s))\\
= &\sup_{0\le u \le s}[H(A(t)) - H(A(u))- \beta(s-u)] \\
& + [H(A(s))\otimes \beta(s) - H(A^*(s))]\\ 
 = &\sup_{0\le u \le s}[H(A(u,t)) - \alpha(t-u) + \alpha (t-u) - \beta(s-u)]\\
 & + [H(A(s))\otimes \beta(s) - H(A^*(s))]\\
 \le & \sup_{0\le u \le s}[H(A(u,t)) - \alpha(t-u)] + \alpha\oslash\beta(t-s) \\
 & + [H(A(s))\otimes \beta(s) - H(A^*(s))]\\ 
\end{align*}
We thus have 
\begin{align*}
&\sup_{0\le s \le t}[H(A^*(s,t)) - \alpha\oslash\beta(t-s)\\
\le & \sup_{0\le u \le t}[H(A(u,t))-\alpha(t-u)]\\
    & + \sup_{0\le s \le t}[H(A(s))\otimes \beta(s) - H(A^*(s))] 
\end{align*}
From the above inequality, the lemma is proved with the definition of information stochastic arrival curve, the definition of information stochastic service curve, Lemma~\ref{lemma:1}, and Lemma~\ref{lemma:2}. \QED
}

\nop{Based on Lemmas~\ref{lemma:clipper} and~\ref{lemma:output}, it is easy to get the following theorem. 

\begin{theorem}\label{theorem:output} \textbf{(Output)} Consider a node $S$ with input flow $A$. If $A \sim_{i.s.a.}<f, \alpha>$, and the node includes an information clipper $C(t)$ and a \textit{i.s.s.} server $S\sim_{i.s.s.}<g,\beta>$, then the output flow $A^*\sim_{i.s.a.}<f\otimes g, (\alpha-H(c))\oslash\beta>$.
\end{theorem}
}

\begin{theorem}\label{theorem:superposition} \textbf{(Superposition)} Consider two flows $A_1(t)$ and $A_2(t)$. Let $A(t)$ denote the aggregate flow, i.e., $A(t)=A_1(t)+A_2(t)$. If for both flows $A_i\sim_{i.s.a.}<f_i, \alpha_i>, i=1, 2,$ and $I(A_1;A_2) \sim_{l.i.s.a.}<\theta,\gamma>,$ then $A\sim_{i.s.a.}<f,\alpha>,$ where $f(x)= (f_1\otimes f_2 \odot \theta)(x)$, and $\alpha(t) = \alpha_1(t) + \alpha_2(t)-\gamma(t)$.    
\end{theorem}
Theorem~\ref{theorem:superposition} means that if two flows follow \textit{i.s.a.} curves, their information fusion (i.e., flow aggregation with information redundancy removed) also follows an \textit{i.s.a.} curve.  

\begin{theorem}\label{theorem:concatenation} \textbf{(Concatenation)} Consider a flow $A(t)$ passing through a network of $N$ nodes in tandem. If each node provides service $S^i \sim_{i.s.s.}<g^i,\beta^i>, i=1,2,\ldots,N$, then the network guarantees to the flow a service $S\sim_{i.s.s.}<g,\beta>$ with 
\begin{align*}
&\beta(t) = \beta^1\otimes\ldots\otimes\beta^N(t)\\
&g(x) = g^1\otimes\ldots\otimes g^N(x).
\end{align*}
\end{theorem}
Theorem~\ref{theorem:concatenation} indicates that the service provided by an end-to-end path follows an \textit{i.s.s.} curve if the nodes along the path follow \textit{i.s.s.} curves. 

\begin{theorem}\label{theorem:output} \textbf{(Output)} Consider a node with input flow $A$. If $A \sim_{i.s.a.}<f, \alpha>$, and the node provides the flow with service $S\sim_{i.s.s.}<g,\beta>$, then the output $A^*\sim_{i.s.a.}<f\otimes g, \alpha\oslash\beta>$.
\end{theorem}
Theorem~\ref{theorem:output} means that the input flow and the output flow both follow \textit{i.s.a.} curves, if the service node follows an \textit{i.s.s.} curve.

\nop{Based on Lemmas~\ref{lemma:clipper} and~\ref{lemma:output}, it is easy to get the following theorem. 
\begin{theorem}\label{theorem:output} \textbf{(Output)} Consider a node $S$ with input flow $A$. If $A \sim_{i.s.a.}<f, \alpha>$, and the node includes an information clipper $C(t)$ and a \textit{i.s.s.} server $S\sim_{i.s.s.}<g,\beta>$, then the output flow $A^*\sim_{i.s.a.}<f\otimes g, (\alpha-H(c))\oslash\beta>.$
\end{theorem}
}

The following theorem illustrates the service guarantee provided by a network node in terms of information backlog, information delay, and information backlog within a delay bound. 
\begin{theorem}\label{theorem:service}\textbf{(Service guarantee)} If the input flow has $A\sim_{i.s.a.}<f,\alpha>$, and the network node provides the flow with service $S\sim_{i.s.s}<g,\beta>$, then 
\begin{enumerate}
\item The information backlog $B(t)$ of the flow at time $t$ satisfies: for all $t\ge 0$ and all $x\ge 0$, 
\begin{equation}
Prob\{B(t)>x\} \le f\otimes g(x- \alpha\oslash\beta(0)).
\end{equation}
\item The information delay $D(t)$ of the flow at time $t$ satisfies: for all $t\ge 0$ and all $x\ge 0$, 
\begin{equation}
Prob\{D(t)>h(\alpha^{x}, [\beta]^{x})\} \le f\otimes g(x),
\end{equation}
where $\alpha^{x}(t) \equiv \alpha(t) + x$, $[\beta]^{x}(t) \equiv \max\{\beta(t), x\}$, and $h(\alpha,\beta)$ is the maximum horizontal distance between functions $\alpha$ and $\beta$ and is defined as 
$$h(\alpha,\beta)= \sup_{s\ge0}\{\inf\{\tau \ge 0: \alpha(s)\le \beta(s+\tau)\}\}.$$ %$\beta'(t) = [\beta(t)-x]^+ + x$
\item The information backlog within delay bound $\tau (\le D(t))$ of the flow at time $t$ satisfies: for all $t\ge 0$ and all $x\ge 0$, 
\begin{equation}
Prob\{\hat{B}(t,\tau)>x\} \le f\otimes g(x+\inf_{v\ge 0}[\beta(v)-\alpha(v-\tau)]).
\end{equation}
Note that $\alpha(t) =0$ for $t\le0$.
\end{enumerate}
\end{theorem}

%% System Modeling
\section{System Model of Information-Driven Networks}\label{sec:model}
\begin{figure}[tp]
 \centerline {\epsfxsize = 3.0 in \epsfbox{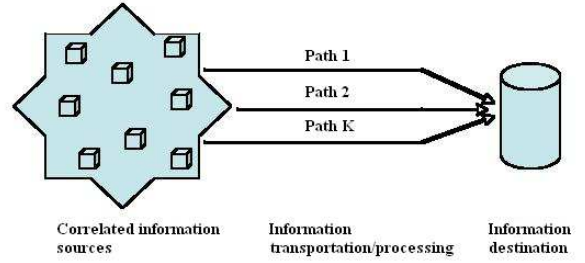}}
  \caption{The architecture of information-driven networks}\label{fig:arc}
\end{figure}

The architecture of information-driven networks is illustrated in Figure~\ref{fig:arc}, including information sources, the transportation paths, and the information destination (or the sink). To simplify presentation, we only consider one sink node. The model, however, could be  easily extended to multiple information sinks. \nop{For ease of reference, the notations used in the paper is listed in Table~\ref{tab:tab1}.}

\subsection{The Model for Information Sources}\label{sec:sourceModel}
Assume that there are $M$ information sources in the system, denoted as $A_1, A_2, \ldots, A_M$, respectively. We assume that an information source $A_i (i=1, \ldots, M)$ has a certain amount of information to send to the destination. We denote the traffic sent from $A_i$ as $A^0_{i}(t)$ and the traffic finally arrives at the destination as $A^{h_i}_{i}(t)$, where $h_i$ is the number of hops of the path used for the information delivery from $A_i$ to the destination. \nop{By default, we use the subscript to denote the number of paths and the superscript to denote the hop count along a path. To simplify notation, we also use $A_i$ to denote the data traffic sent from information source $A_i$. }

\nop{In many applications, such as sensor monitoring systems or peer-to-peer content streaming, data from the same source usually exhibit temporal correlation, and as such from the same source, the loss of some data packets may not result 
in a large loss in information. Formally, for the information from source $S_i$ to the destination over a lossy path, we require that 
\begin{equation}
I(A^0_i(t); A^{h_i}_{i}(t+\tau)) > \alpha H(A^0_i(t)),\label{eq:2}
\end{equation}
where $\alpha$ is a given positive threshold value no larger than $1$ and $\tau$ is the end-to-end delay described in Section~\ref{sec:background}. \nop{Eqution~\ref{eq:2} is to model the temporal correlation in an information source. }   
}

For the $M$ information sources, we use $I(A^0_i(t);\ldots,A^0_j(t))$, $1\le i < j \le M$  to model the information redundancy among different data sources. 

\subsection{The Model for Transportation Paths} \label{sec:networkModel}
We assume that $K$ paths, denoted as $L_1, \ldots, L_K$,  from the set of information sources to the destination are used for information delivery. In this paper, we only focus on $K$ paths that are node-disjointed. When multiple paths share some common intermediate nodes, the number of possible schemes for information fusion becomes tremendous, in which case it is extremely difficult to derive the performance bounds. We leave the study of multiple paths sharing intermediate nodes as our future work.  Nevertheless, we do not assume the data transmissions along the $K$ paths are always independent. For example, in wireless networks even if two paths do not share common intermediate nodes, the transmissions along the two paths may interfere with each other. In short, we assume that the transmissions along different paths may impact each other, but information fusion is only performed at the entrance nodes and the sink node. As a reminder, the information service in Definition~\ref{def:service} includes both information processing and information transmission. 

 \nop{When the transmissions along two paths do not impact each other, we call the two paths independent, otherwise they are called correlated paths. } 

In addition, we assume that an end-to-end path, once established, is fixed for the time period of interest. Networks in which an end-to-end path changes quickly over time are beyond the focus of this paper and are left as our future work.

We denote a path $L_i$ of length $h_i$ (i.e., the hop count from the source to the destination is $h_i$) as $L_i=(L_i^0, L_i^1, \ldots, L_i^{h_i})$, where $L_i^j$ denotes the $j$-th node of path $i$. Each node $L_i^j$ provides an information stochastic service curve $S \sim_{i.s.s}<g_i^j, \beta_i^j>$ along the path. 

When two paths, $L_i= (L_i^0, L_i^1, \ldots, L_i^{h_i})$ and $L_j= (L_j^0, L_j^1, \\ \ldots, L_j^{h_j})$ interfere with each other in data transmission, we define an $h_i \times h_j$ matrix $[\hat{\mathcal{I}}_{(l,m)}], l=0, \ldots, h_i-1, m=0, \ldots, h_j-1$ to capture the transmission correlation along the two paths, where $\hat{\mathcal{I}}_{(l,m)}$ is an information impairment process that should be integrated in the information service curves of the nodes $L_i^l$ and $L_j^m$. This information impairment process is introduced to reduce the information service rate along the paths involved. Note that  $L_i^{h_i} =  L_j^{h_j}$ since we assume the same destination. 

To investigate the impact of transmission interference, the following theorem will be used to adjust the service rate of an impacted node. Its proof is provided in Appendix~\ref{app:A}. 
\begin{theorem}\label{theorem:impair} \textbf{(Service reduction with information impairment)} Consider a network node providing a flow with service $S\sim_{i.s.s.}<g,\beta>$. If the node is interfered with an impairment process $\hat{\mathcal{I}}$ with information stochastic arrival curve $\hat{\mathcal{I}}\sim_{i.s.a.}<f,\alpha>$, then the network node guarantees to the flow a service $S\sim_{i.s.s.}<g\otimes f, \beta-\alpha>$.  
\end{theorem}

\subsection{Problems of Interest}\label{sec:problems}
Thus far, we have presented a calculus and an abstract model for information-driven networks. Among many interesting problems in this analytical framework, we are interested in the following key questions: \nop{ for a system with a set of information sources $A=<A_1, \ldots, A_M>$ and a set of available transportation paths $P=<P_1, \ldots, P_K>$:}
\begin{enumerate}  
\item What are the stochastically achievable information delivery rates of a network? The answer to the question discloses the limit of information delivery in the system.
\item Given a set of information sources and an end-to-end delay bound, can the total information be delivered to the destination with a high probability, and how?  The answer to the question provides insights on the design of scheduling algorithm for information-driven networks. 
\nop{\item When end-to-end delay is considered, how much information could be delivered to the destination within the bound? what is the optimal transmission schedule so that the information delivery ratio is maximized within a delay bound? The answer to the question provides insights on the design of scheduling algorithms for information-driven networks. 
Given a desired information delivery ratio, what is the optimal transmission schedule so that the total amount of data transmitted is minimized? The answer to the question discloses the bandwidth saving when the network becomes information driven. }    
\end{enumerate}

In this paper, we focus only on the feasibility of information delivery and performance bounds, but intentionally avoid the study of many interesting optimization problems within the analytical framework. This is because the main purpose of this paper is to present a new analytical approach and its application for information-driven networks and also because traditional measurements on resource overhead have not been well defined in information-driven networks. For instance, we need to mathematically formulate the computational overhead for information processing. We leave these aspects as our future work. 

%% Analysis for Ideal case
\section{Stochastically Achievable Information Delivery}\label{sec:analysis}
\begin{figure}[tp]
 \centerline {\epsfxsize = 3.0 in \epsfbox{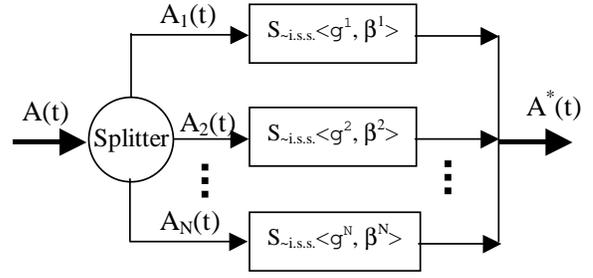}}
  \caption{An information splitter for servers in parallel}\label{fig:parallel}
\end{figure}

We answer the first question in this section. For this, we need to investigate the information service guarantee provided by multiple servers in parallel, as shown in Figure~\ref{fig:parallel}. For simplicity, we assume a fluid model in which information could be split in infinitesimal amounts. This constraint will be relaxed in the next section. 

\begin{definition} \textit{A weighted information splitter} is a scheduler that splits an input flow $A(t)$ into multiple \textit{information exclusive} sub-flows $A_1(t), \ldots, A_N(t)$, with each assigned a weight $w_i$ and served by an information processing node $S^i (i=1, \ldots, N)$, respectively. At any time instant $t$, the sub-flow assigned to $S^i$ satisfies 
$H(A_i(t)) = \frac{w_i}{\sum_{j=1}^{N} w_j} H(A(t))$. 
 \nop{with weight $w_i$ assigned to sub-flow $A_i(t), 1\le i \le N$, such that $H(A(t)) = H(A_1(t)) + \ldots, H(A_N(t))$ and for $\forall M \le N$, $I(A_1(t); \ldots; A_M(t)) =0$. Equivalently, at any time instant $t$, $H(A_i(t)) = \frac{w_i}{\sum_{j=1}^Nw_j} H(A(t))$ holds for any $1\le i \le N$.}
\end{definition}
\begin{theorem}\label{theorem:parallel} \textbf{(Parallel servers)} Consider a flow $A(t)$ passing through a weighted information splitter and then a network of $N$ nodes in parallel (Figure~\ref{fig:parallel}). Assume that all nodes are work conserving, i.e., they cannot become idle if there is information waiting for service. Assume that each node provides service $S^i \sim_{i.s.s.}<g^i,\beta^i> (i=1,2,\ldots,N)$ and the weight of the sub-flow to $S^i$ at time $t$ is set to $\beta^i(t)$. \nop{Assume that the flow $A(t)$ arrives at an information rate no smaller than $(\beta^1+\ldots+\beta^N)(t)$, i.e., $H(A(t)) \ge (\beta^1+\ldots+\beta^N)(t)$. }The whole system guarantees to the flow a service $S\sim_{i.s.s.}<g,\beta>$ with 
\begin{align*}
&\beta(t) = (\beta^1+\ldots+\beta^N)(t)\\
&g(x) = g^1\otimes\ldots\otimes g^N(x).
\end{align*}
\end{theorem}

It is worth highlighting that Theorem~\ref{theorem:parallel} is intended to disclose a network's achievable information delivery rate, and for this reason it assumes an ideal situation in which information could be split in infinitesimal amounts. In practice, however, such an assumption may not hold and the network's actual information delivery rate would be lower.  

Equipped with the theorems in this paper, Algorithm RateCal, shown in Figure~\ref{fig:alg}, can be used to search for stochastically achievable information delivery rates of a network.  It uses a brutal-force search and has a time complexity of $O(2^K)$, where $K$ is the number of node-disjoint paths from the sources to the destination. The algorithm is not scalable with $K$, but practically $K$ is usually not very large. In addition, it is very easy to reduce the complexity by considering the stochastic relationship of \textit{i.s.s.} curves. A curve $S^1\sim_{i.s.s.}<g^1,\beta^1>$ is considered better than another curve $S^2\sim_{i.s.s.}<g^2,\beta^2>$ \textit{iff} for $\forall t, x \ge 0$, $\beta^1(t) > \beta^2(t)$ and $g^1(x) \le g^2(x)$. In this case, we can ignore the curve $S^2$ to reduce the search space.  

\begin{figure}[t]
\setcounter{line}{0}
\begin{tabbing}
1234\=56\=78\=90\=01\=23\=45 \kill
  {\bf Input:} A network with $K$ parallel paths (Figure~\ref{fig:parallel});\\
  \>  \textit{i.s.s.} curve of each node;\\
  \> A set of impairment matrices; \\
  {\bf Output:} a list of stochastically achievable services;\\
  {\bf Method:}\\
  \addtocounter{line}{1}\theline:
  {\tt FOR} any subset of the $K$ paths, $\{L_i, \ldots, L_j\}$, \\
  \>where $1\le i\le j \le K$ $\{$ \\
  \addtocounter{line}{1}\theline:
  \> {\tt FOR} $n = i$ {\tt TO} $j \{$\\
  \addtocounter{line}{1}\theline:
  \>\> {\tt FOR} each node along path $L_n$, if its transmission\\
  \>\>  is interfered by another path in the subset, \\ 
    \addtocounter{line}{1}\theline:
 \>\>\> Adjust the node's \textit{i.s.s.} with Theorem~\ref{theorem:impair} and \\
  \>\>\>the \textit{i.s.a.} curves in relevant impairment matrices; \\
      \addtocounter{line}{1}\theline:
  \>\> Calculate the \textit{i.s.s.} of path $L_n$ with Theorem~\ref{theorem:concatenation};\\
  \>\> $\}$\\
  \addtocounter{line}{1}\theline:
  \> Output the \textit{i.s.s.} of subset $\{L_i, \ldots, L_j\}$ with Theorem~\ref{theorem:parallel};\\
  \>$\}$
 \end{tabbing}
\hrule\caption{\label{fig:alg}RateCal: An algorithm to calculate stochastically achievable information delivery rates.}
\end{figure}

\begin{example} We use a simple example to illustrate the calculation of stochastically achievable information delivery rates of  a network. Assume that a network has three parallel end-to-end paths,  $L_i, i=1, 2, 3$. Assume that path $L_i$ includes $i (=1,2, 3)$ nodes, respectively. Assume that the transmissions along different paths do not interfere with each other. Also assume that all nodes provide the same information stochastic service following $\sim_{i.s.s.}<e^{-x}, rt>$.  Running Algorithm RateCal, we obtain seven \textit{i.s.s.} curves, as shown in Figure~\ref{fig:performance1}. For instance, based on the \textit{i.s.s.} curve $<6e^{-x/6}, 3rt>$, we can see that for a (deterministic) information arrival following $\sim_{i.s.a.}<0, 3rt>$, the network can guarantee its delivery such that up to time $t$, the probability that the output information is less than $3rt - 24$ is smaller than $0.1 (\approx  6e^{\frac{-24}{6}})$.   
\begin{figure}[tp]
 \centerline {\epsfxsize = 2.5 in \epsfbox{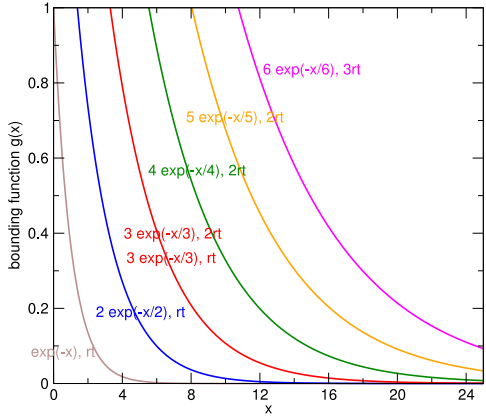}}
  \caption{Stochastically achievable service curves in the example network}\label{fig:performance1}
\end{figure}
\end{example}

\section{Transmission Scheduling}\label{sec:scheduling}
Although the previous section provides the answer to stochastically achievable information delivery rates, the results are based on an ideal assumption, that is, the information could be split in infinitesimal amounts. In practice, however, we are often met with the situation that information cannot be split arbitrarily. In this sense, the results of previous section only provide theoretical upper bounds on stochastic information delivery rates, which may not be practically feasible for a given request. We hence need to study the following problem: Given a set of information sources and an end-to-end delay bound, can the total information be delivered to the destination with a high probability (e.g. no lesser than $p$)? If yes, how? if not, what is the percentage of the information that could be delivered and in what probability?  \nop{As such, we only use the results of previous section as good heuristics in the search for feasible transmission schedules. }   

\nop{In this paper, we use the \textit{information delivery ratio} to capture the network capability in information delivery. The following definitions are necessary:}
\begin{definition} \label{def:schedule}
\textit{A transmission schedule} for a set of information sources $A=<A_1, \ldots, A_M>$ over a set of paths $L=<L_1, \ldots, L_K>$ is denoted as a tuple $<A, L, \pi>$, where $\pi$ is a mapping function from $A$ to $L$, and $\pi(A_i) = L_j$ means that data from source $A_i (1\le i \le M)$ is sent over path $L_j (1\le j \le K)$ in the transmission schedule.  
\end{definition}

\begin{definition} \label{def:ratio}
\textit{Information delivery ratio within delay $\tau$:} For a set of information sources $A = (A_1, \ldots, A_M)$ sent from $M$ sources to a destination, the information delivery ratio at time $t$, $\mathcal{R}(t)$, is defined as
\begin{equation}
\mathcal{R}(t) \equiv \frac{H(A^*(t+\tau))}{H(A(t))} = \frac{H(A(t))-\hat{B}(t,\tau)}{H(A(t))}, \label{eq:R}
\end{equation}
where $A^*(t+\tau)$ represents the set of traffic that has arrived at the information sink at time $t+\tau$. \end{definition}

The scrutinizing readers may have realized that $H(A^*(t+\tau))$ is not well defined because to get it needs calculation of $I(A^*_1; \ldots; A^*_M)$, which has not been defined. The following theorem discloses the relationship of information redundancy before and after information delivery and can be used to overcome this difficulty. 
\begin{theorem}~\label{theorem:correlation}
Consider a set of information sources $A = (A_1, \ldots, A_M)$. Split the set into two subsets $\Delta_1= (A_1, \ldots, A_i)$ and $\Delta_2= (A_{i+1}, \ldots, A_M)$, and transmit $\Delta_1$ and $\Delta_2$ simultaneously along two separate paths at time $t$. Assume that at time $t+\tau$, both $H(\Delta_1)$ and $H(\Delta_2)$ have been received, i.e., $H(\Delta^*_1(t+\tau)) = H(\Delta_1(t))$ and $H(\Delta^*_2(t+\tau)) = H(\Delta_2(t))$. We have $ I(\Delta^*_1(t+\tau); \Delta^*_2(t+\tau)) = I(\Delta_1(t); \Delta_2(t)).$
\end{theorem}
 
Note that Theorem~\ref{theorem:correlation} holds no matter whether the information redundancy \textit{within each set} $\Delta_i, i=1, 2$ is removed or not during the transmission.

With Definitions~\ref{def:schedule} and~\ref{def:ratio}, the formal formulation of the problem becomes: Given a set of information sources $A=(A_1, \ldots, A_M)$, a set of separate transportation paths $L_1, \ldots, L_K$, an end-to-end delay bound $D$, and a probability $p$, is there a transmission schedule to deliver the information from the sources to the destination within time delay $D$ with a probability no smaller than $p$? It is easy to see that this problem is NP-hard even without considering the information redundancy among the information sources, because the problem can be transformed into a ``bin-packing" problem~\cite{Gar}, which is NP-hard. We hence propose a method, called best-fit-largest-redundancy(BFLR) algorithm, to search for an approximate solution. An error may occur if the algorithm returns a negative answer to the existence of a feasible transmission schedule, but a feasible transmission schedule does actually exist. Nevertheless, due to the NP-hardness of the problem, such an error cannot be avoided unless both $K$ and $M$ are very small so that we can use brutal-force search. Our later experimental study shows that BFLR can find almost all feasible transmission schedules.  

\nop{ Assume that $H(A_i(t)), i=1, \ldots, M$ follows  $\sim_{i.s.a.}<f^i,\alpha^i>$ and $H(A(t))$ follows $\sim_{i.s.a.}<f,\alpha>$ (obtained with Theorem~\ref{theorem:concatenation}). The algorithm includes four basic steps: 
\begin{enumerate} 
\item Call Algorithm RateCal to calculate all stochastically achievable service rates, denoted as $\sim_{i.s.s.}<g^i,\beta^i>$, respectively.  Note that a feasible service rate represents a subset of transportation paths.    
\item Sort the service rates in the decreasing order of rate $\beta^i$. Among the stochastically achievable service rates, only the rates that are larger than $\alpha$ will be considered, and we all these rates feasible service rates. 
\item Sort $A_i$ in the decreasing order of information rates $\alpha^i$. 
\item Select the first feasible service rate and equivalently the corresponding subset of paths. Combine information sources that have the largest information redundancy and can best fit to a transportation path. If no fitting path could be found, consider the next feasible service rate and so on. BFLR returns no if all feasible service rates are processed but no feasible transmission schedule could be found.  
\end{enumerate}
}
\begin{figure*}[ht]
\setcounter{line}{0}
\begin{tabbing}
1234\=56\=78\=90\=01\=23\=45 \kill
  {\bf Input:} A network with $K$ parallel paths (Figure~\ref{fig:parallel}), each node's service curve, and impairment matrices;\\
  \>\> A set of information sources $A=(A_1, \ldots, A_M)$ and related information redundancy;\\
  \>\> End-to-end delay bound $D$;\\
  \>\> The violation probability $p$;\\
  {\bf Output:} Feasible transmission schedules, or no feasible transmission schedule;\\
  {\bf Method:}\\
  \addtocounter{line}{1}\theline: 
Call Algorithm RateCal to obtain all stochastically achievable service rates, each characterized by $\sim_{i.s.s.}<g^i,\beta^i>$; \\
    \addtocounter{line}{1}\theline:
Remove the achievable service rates if its rate is smaller than the information arrival rate of $H(A)$;\\
    \addtocounter{line}{1}\theline:
Sort the remaining achievable service rates in the decreasing order of rate;\\ 
     \addtocounter{line}{1}\theline:
 Sort $A_i$ in the decreasing order of their information arrival rates; //Note: $A_i\sim_{i.s.a.}<f^i, \alpha^i>$.\\   
     \addtocounter{line}{1}\theline:
{\tt FOR} each remaining achievable service rate in the sorted order $\{$ \\
     \addtocounter{line}{1}\theline:
     \>  Find the subset of paths corresponding to the current achievable service rate; \\ 
     \>\> // Note: A feasible service rate represents a subset of paths. \\
     \addtocounter{line}{1}\theline:
     \> Sort the paths in the decreasing order of their service rates;\\
          \addtocounter{line}{1}\theline:
     \> $AS = \emptyset$; // Note: $AS$ is an initially empty set to store information sources that could be combined together. \\
     \addtocounter{line}{1}\theline: 
     \>  {\tt FOR} each path in the subset $\{$ \\ 
          \addtocounter{line}{1}\theline:
     \>\> Find the information source whose information arrival rate best fits the current path; \\
     \>\>\> // Note:  Best-fit means that the source's information arrival rate is smaller than the service rate of the path, \\
     \>\>\>\> but is the largest among all non-processed information sources.\\
                   \addtocounter{line}{1}\theline:
     \>\> Use Theorem~\ref{theorem:service}.(2) to check if the information source can be delivered with the current path to meet the requirement;\\
           \addtocounter{line}{1}\theline:    
     \>\>Go to the next path if best-fit cannot be found;\\
           \addtocounter{line}{1}\theline:
      \>\> Label the best-fit information source as \textit{processed} and insert it into $AS$;\\
      \>\> {\tt REPEAT} $\{$\\
           \addtocounter{line}{1}\theline:         
      \>\>\> Find the non-processed information source that has the largest redundancy with $AS$;\\
                \addtocounter{line}{1}\theline:
        \>\> $\}$ {\tt UNTIL} $AS$ cannot be delivered with the current path to meet the delay requirement; \\
\addtocounter{line}{1}\theline:
        \>\> {\tt IF} all information sources have been processed, output the transmission schedule; \\
        \>\>\> {\tt ELSE} Go to the next path. \\
      \>$\}$\\
      \addtocounter{line}{1}\theline:
      \> Go to the next achievable service rate;\\   
      \addtocounter{line}{1}\theline: 
      \> Return NO if all achievable service rates have been checked but no feasible transmission schedule has been found.\\
      \addtocounter{line}{1}\theline:
     $\}$
 \end{tabbing}
\hrule\caption{\label{alg:BFLR}BFLR: An algorithm to search for feasible transmission schedules.}
\end{figure*}

The basic idea of the BFLR algorithm is to use the stochastically achievable service rates obtained by Algorithm RateCal as the guideline in the selection of a feasible subset of paths, and combine information sources that have the largest information redundancy and can best fit to a transportation path. The detailed steps of BFLR is illustrated in Figure~\ref{alg:BFLR}.   

When the BFLR algorithm returns no, it may be because either (1) the information service rate of a selected path is smaller than the combined rate of a subset of sources, or (2) the information service rate of a selected path is larger than the combined rate of a subset of sources but the information cannot be delivery in the (tight) delay bound. In the first case, no feasible transmission schedule exists even if we enlarge the delay bound. In the latter case, the problem that we need to answer is: what is the information delivery ratio $\mathcal{R}$ within the delay bound $D$ and in what probability? The problem can be answered by revising the BFLR algorithm and using Theorem~\ref{theorem:service}.(3). Basically, we should remove Line 18 of the algorithm, and in Line 11 use Theorem~\ref{theorem:service}.(3) to check the information backlog within the delay bound and the corresponding stochastic bounding function. In addition, the output (Line 16) should be changed accordingly. We omit the details to avoid triviality.

\nop{
The above definition is quite general and applicable for different networking scenarios. For a ``never-stop" streaming system, e.g., a wireless sensor network that continuously sends out measurement data, we normally set $\tau \le \min\{D_1(t),\ldots, D_K(t)\}$ where $D_i(t)$ denote the information delay along path $L_i, i=1, \ldots, K,$ to investigate information delivery ratio within a time window; for a ``one-time" information system, e.g., a distributed peer-to-peer file sharing system where traffic from a source terminates after transmission, we can set $\tau \rightarrow \infty$ to evaluate how much information can be transported.   }

\section{Case Study}\label{sec:evaluation}

\subsection{Network Configuration}
We use an exemplary information-driven network to illustrate how the results in this paper can be used for performance evaluation in practice. 

\textbf{Information sources.} We consider a network consisting of $M=9$ information sources. These sources
are partitioned into 3 groups with sources $1\sim 3$ in Group 1, sources $4\sim 6$
in Group 2, and sources $7 \sim 9$ in Group 3. Information from the same group exhibits spatial correlation, but information from different groups is assumed to be independent.  

\textbf{Temporal correlation.} Assume that the information of each
source is collected by periodically
sampling a stationary Gaussian stochastic process. Concretizing the
information
$H(\cdot)$ of the flow $A_i$ generated by source $i$ to the Shannon entropy
function~\footnote{Shannon's entropy function is consistent with our
definition of information in Section~\ref{sec:informationNotation}},
it yields for discrete time $t$~\cite{Cov}:
\begin{eqnarray}
H(A_i(t)) = \alpha_i(t)=\frac{1}{2}\log (2 \pi e)^t
|\mathcal{C}_i^{(t)}|, \hspace{0.2cm} t=1,2,...
\label{equ:tem_correlation}
\end{eqnarray}
where $\mathcal{C}_i^{(t)}$ is the $t\times t$ covariance matrix for the flow of
source $i$ and
is specified by the temporal covariance function $\Gamma_i(\tau)$,
i.e., the matrix element $\mathcal{C}_i^{(t)}(j,k)=\Gamma_i(k-j)$, where $1\le j, k \le t$.
Here, we adopt the typical exponential covariance function~\cite{Vur}:
\begin{eqnarray}
\Gamma_i(\tau)=\sigma_i^2 e^{-|\tau|/\eta_i}, \hspace{0.2cm}
\tau=0,\pm 1,\pm 2,...
\label{equ:covariance_func}
\end{eqnarray}
where $\sigma_i^2$ is the variance of flow $A_i$ and $\eta_i$ is a constant.
Assume that each source generates messages at a constant interval $\delta$.
By applying Equation~(\ref{equ:covariance_func}) in
(\ref{equ:tem_correlation}),
the stochastic arrival curve of flow $A_i$ can be specified in the continuous time $t$
as $A_i \sim_{i.s.a.} <0,\alpha_i>$, where\\
\begin{eqnarray}
\label{equ:single_arr_curve}
&& \alpha_i(t)=  \\ \nonumber
&&\begin{cases}
\frac{t}{2 \delta} \log(2 \pi e \sigma_i^2), 0\le t \le \delta \\
\frac{t}{2 \delta} \log(2 \pi e \sigma_i^2
(1-e^{-2/\eta_i}))-\frac{1}{2}\log(1-e^{-2/\eta_i}), t>\delta
\end{cases}
\end{eqnarray}

\nop{
where $\sigma_i^2$ is the variance of flow $i$ and $\eta_i$ is a constant.
Thus, the \textit{i.s.a.} curve of flow $A_i$ can be specified
as $A_i \sim_{i.s.a.} <0,\alpha_i>$. where
\begin{eqnarray}
\alpha_i(t) = \frac{t}{2} \log(2 \pi e \sigma_i^2 (1-e^{-2/\eta_i}))-\frac{1}{2}\log(1-e^{-2/\eta_i})
\end{eqnarray}

In this study, we set $\sigma_i^2=1$ and $\eta_i=10$ for all sources.}

\textbf{Spatial correlation.} In the same source group, we model the information redundancy of the sources using a spatial correlation model similar to that in~\cite{Vur}. Specifically, we assume that for sources in the same group,
\begin{eqnarray}
H(\sum_i A_i) = \sum_i \epsilon_i H(A_i), \hspace{0.2cm} 0 \le \epsilon_i \le 1 \nonumber
\label{equ:spa_correlation}
\end{eqnarray}

where $\epsilon_i$ are constants and depend on the sources' locations and the adopted spatial model~\cite{Vur}. In our study, we use the same setting of information redundancy for all three groups. Specifically, for the three sources in a group, denoted as $A_1,A_2,A_3$, we set:
\begin{eqnarray}
& &H(A_1+A_2) = H(A_2+A_3)= H(A_1+A_3)= 1.8 H(A_1) \nonumber \\
& & H(A_1+A_2+A_3)= 2.4 H(A_1)\nonumber
\end{eqnarray}
While the above parameter settings are kind of arbitrary, we stress that \textit{using other parameters has no impact on the effectiveness of our analytical framework} since users can adopt any other temporal and spatial correlation models different from our example scenario in their analysis.

\nop{
\textbf{Information arrival rate.} Given temporal and spatial correlation models, we can determine the information arrival rate by setting a time interval. Note that in Equation~\ref{equ:tem_correlation}, $t$ is a time index which does not necessarily mean the actual time.  Assume that $\eta_i=100$. We set the parameter $\sigma_i$ for each source $A_i \sim_{i.v.b.} <0,\alpha_i>$ such that 
$\alpha_i(t) \sim \frac{7}{24}rt$. Assume that each source generates messages at a constant interval of $100$ ms, i.e., the time index increases by one every $100$ ms.
}
\begin{figure}[ht]
\centering
\includegraphics[width=0.4\textwidth]{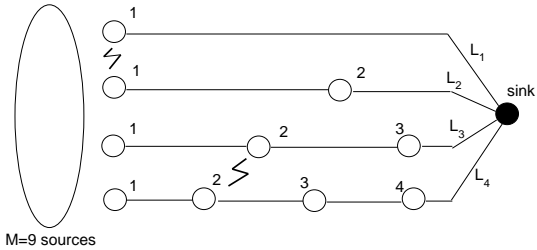}
\caption{An example network} 
\label{fig:example}
\end{figure}

\textbf{Transportation paths.} As shown in Fig.~\ref{fig:example}, the collected information is transmitted to a single information sink through  
four parallel end-to-end paths, $L_i,i=1,2,3,4$. 
On each path $L_i$, there are $i$ nodes. Assume that all network nodes provide information service following $\sim_{i.s.s.}<e^{-x},r(t-d)>$, where
$r$ is the average information service rate and $d$ is the per-hop delay. Also assume that 
the first node in $L_1$ and the first node in $L_2$ are subject to correlated impairment, which follows  $\sim_{i.s.a.}<4 e^{-x/4}, r(t-d)/5>$. Similarly, assume that 
the second node in $L_3$ and the second node in $L_4$ are subject to correlated impairment, which follows  $\sim_{i.s.a.}<3 e^{-x/3}, r(t-d)/3>$. 

Note that Example~\ref{ex:motivation} is a simplified special case of this case study.

\subsection{Numerical Results}
We would like to know whether the information network could deliver the information from sources to the destination within a given delay bound with at least a certain probability. If feasible transmission schedules cannot be found, we also want to know the information delivery ratio with at least a certain probability. Since the performance measures, e.g., delay and information delivery ratio, are time dependent. we calculate their maximum value to remove the time dependancy in this case study. For example, the delay is calculated as $\sup_{t\ge0} D(t)$.     

In the example study, we select the rate parameter $r=8$~kbps, per-hop delay $d=7.5$~ms. For every information source, we set $\delta=100$~ms and $\eta_i=100$. With reference to Equation~(\ref{equ:single_arr_curve}),  $\sigma_i$ is set such that the long-term information rate of a source $ \approx 2.33$~kbps. Considering the spatial correlation modeled, the total long-term information arrival rate of the $9$ sources amounts to $16.78$~kbps. 

First, we calculate the information stochastic service curve of each path and list the results in Table~\ref{tab:1}.
\begin{table}
\caption{Information stochastic service curve of each path}\label{tab:1}
\begin{tabular}{|c|c|c|}\hline
Path  & w/o impairment & w/ impairment\\
\hline
\hline
$L_1$ & $<e^{-x},r(t-d)>$ & $<5e^{-x/5}, \frac{4}{5}rt-\frac{4}{5}rd)>$ \\
\hline
$L_2$ & $<2e^{-x/2},r(t-2d)>$ & $<6e^{-x/6}, \frac{4}{5}rt-\frac{9}{5}rd)>$ \\
\hline
$L_3$ & $<3e^{-x/3},r(t-3d)>$ & $<5e^{-x/5}, \frac{2}{3}rt-\frac{8}{3}rd)>$ \\
\hline
$L_4$ & $<4e^{-x/4},r(t-4d)>$ & $<6e^{-x/6}, \frac{2}{3}rt-\frac{11}{3}rd)>$ \\
\hline
\end{tabular}
\end{table}

To facilitate understanding, we list the subsets of paths found by Algorithm BFLR that have total information service rate larger than the total information arrival rate, i.e., the intermediate result after Step 2 of Algorithm BFLR:
\begin{enumerate}
\item $L_1+L_2+L_3$ $\sim_{i.s.c.} <14 e^{-t/14} , \frac{13}{5}rt-\frac{28}{5}rd>$
\item $L_1+L_2+L_4$ $\sim_{i.s.c.} <15e^{-t/15}, \frac{13}{5}rt-\frac{33}{5}rd>$
\item $L_1+L_3+L_4$ $\sim_{i.s.c.} <12 e^{-t/12}, \frac{7}{3}rt-\frac{22}{3}rd>$
\item $L_2+L_3+L_4$ $\sim_{i.s.c.} <13 e^{-t/13} , \frac{7}{3}rt-\frac{25}{3}rd>$
\item $L_1+L_2+L_3+L_4$ $\sim_{i.s.c.} <22 e^{-t/22} , \frac{44}{15}rt-\frac{134}{15}rd>$
\end{enumerate}
The first item means that if path $1$, path $2$, and path $3$ are used, the achievable stochastic information service follows $\sim_{i.s.s.} <14 e^{-t/14} , \frac{13}{5}rt-\frac{28}{5}rd>$. The explanation of other items is similar.

We can search for feasible transmission schedules using Algorithm BFLR for different 
delay bounds (e.g. $35$ ms and $45$ ms) and for different violation probabilities
(e.g., $Prob=0.001$ and $Prob=0.0001$). The results are shown in Table~\ref{tab:2}. 
In the table, ``X'' denotes that the correspondent path set is not feasible. ``Ai.j'' stands for the source $j$ in Group $i$. As an example, the first three lines on the third column mean that if we send information from sources $A1.1, A1.2, A1.3$ along path $L_1$, information from sources $A2.1, A2.2, A2.3$ along path $L_2$, and information from sources $A3.1, A3.2, A3.3$ along path $L_3$, the total information can be received within delay bound $35$ ms with a violation probability no larger than $0.001$. Because information from the same source group exhibits the largest redundancy, the results demonstrate that Algorithm BFLR is capable of finding a transmission schedule so that information from the same source group is fused together to remove information redundancy.

\begin{table}[ht]
\caption{Results of BFLR: feasible transmission schedules} \label{tab:2}
\begin{tabular}
{|l|l||l|l|}\hline
 \multicolumn{2}{|c||}{$Prob$ /  Delay bound} & 35 ms & 45 ms\\
\hline
\hline
 & & $L_1:A1.1 \sim 1.3$& $L_1: A1.1 \sim 1.3$ \\ \cline{3-4}
& ($L_1,L_2,L_3$) & $L_2: A2.1 \sim 2.3$ & $L_2: A2.1 \sim 2.3$ \\ \cline{3-4}
&& $L_3: A3.1 \sim 3.3$ & $L_3: A3.1 \sim 3.3$ \\ \cline{2-4}
 & & $L_1: A1.1 \sim 1.3$ & $L_1: A1.1 \sim 1.3$ \\ \cline{3-4}
& ($L_1,L_2,L_4$) & $L_2: A2.1 \sim 2.3$ & $L_2: A2.1 \sim 2.3$ \\ \cline{3-4}
.1\%&& $L_4: A3.1 \sim 3.3$ & $L_4: A3.1 \sim 3.3$ \\ \cline{2-4}
& ($L_1,L_3,L_4$) & X & X \\ \cline{2-4}
& ($L_2,L_3,L_4$) & X & X \\ \cline{2-4}
 & & & $L_1: A1.1 \sim 1.3$  \\ \cline{4-4}
& ($L_1,L_2,L_3,L_4$) & X & $L_2: A2.1 \sim 2.3$ \\ \cline{4-4}
& & & $L_3: A3.1 \sim 3.2$ \\ \cline{4-4}
&& & $L_4: A3.3$  \\ 
\hline

 & & $L_1:A1.1 \sim 1.3$& $L_1: A1.1 \sim 1.3$ \\ \cline{3-4}
& ($L_1,L_2,L_3$) & $L_2: A2.1 \sim 2.3$ & $L_2: A2.1 \sim 2.3$ \\ \cline{3-4}
&& $L_3: A3.1 \sim 3.3$ & $L_3: A3.1 \sim 3.3$ \\ \cline{2-4}
 & & & $L_1: A1.1 \sim 1.3$ \\ \cline{4-4}
& ($L_1,L_2,L_4$) & X & $L_2: A2.1 \sim 2.3$ \\ \cline{4-4}
.01\%&& & $L_4: A3.1 \sim 3.3$ \\ \cline{2-4}
& ($L_1,L_3,L_4$) & X & X \\ \cline{2-4}
& ($L_2,L_3,L_4$) & X & X \\ \cline{2-4}
 & & & $L_1: A1.1 \sim 1.3$  \\ \cline{4-4}
& ($L_1,L_2,L_3,L_4$) & X & $L_2: A2.1 \sim 2.3$ \\ \cline{4-4}
& & & $L_3: A3.1 \sim 3.2$ \\ \cline{4-4}
&& & $L_4: A3.3$  \\ 
\hline
\end{tabular}
\end{table}

As discussed before, when the BFLR algorithm returns no, it may be because either (1) the information service rate of a selected path is smaller than the combined rate of a subset of sources, or (2) the information service rate of a selected path is larger than the combined rate of a subset of sources but the information cannot be delivery within the delay bound. We find that the two subsets of paths, $<L_1, L_3, L_4>$ and $<L_2, L_3, L_4>$, belong to the former case, meaning that no feasible transmission schedule exists even if we enlarge the delay bound. The other ``X"s in Table~\ref{tab:2} belong to the latter case, where we want to know the information delivery ratio within the delay bound. 

By setting even tighter delay bounds, $15$ ms and $20$ ms for example, we can make the rest three subsets of paths in Table~\ref{tab:2} belong to the latter case. The results regarding information delivery ratio are shown in Table~\ref{tab:3}. Since the results are calculated using Theorem~\ref{theorem:service}.(3) and Equation~(\ref{eq:R}), the results are explained in the following way: the probability that the information delivery ratio within a delay bound is smaller than a value is not larger than a violation probability. For instance, 
the first line of the third column indicates that using paths $L_1, L_2$, and $L_3$, the probability that the information delivery ratio within the delay bound of $15$ ms is smaller than $56.4\%$ is not larger than $0.1$. The ``twist" logic implies that using paths $L_1, L_2$, and $L_3$, with a probability no less than $0.9$, the information delivery ratio within the delay bound of $15$ ms is at least $56.4\%$.      

\begin{table}[ht]
\centering
\caption{Information delivery ratio under small delay constraints}~\label{tab:3} % under small delay constraints
\begin{tabular}{|l|l||l|l|}\hline
\multicolumn{2}{|c||}{$Prob$ /Delay bound} & 15 ms & 20 ms\\
\hline
\hline
& ($L_1,L_2,L_3$) & 56.4\%& 63.8\% \\ \cline{2-4}
0.1& ($L_1,L_2,L_4$) & 50.6\%& 56.1\% \\ \cline{2-4}
& ($L_1,L_2,L_3,L_4$) & 50.2\% & 57.8\% \\ \cline{4-4}
\hline

& ($L_1,L_2,L_3$) & 59.7\% & 67.1\%\\ \cline{2-4}
0.15& ($L_1,L_2,L_4$) & 53.9\% & 59.5\% \\ \cline{2-4}
& ($L_1,L_2,L_3,L_4$) & 54.2\% & 61.8\% \\ \cline{4-4}
\hline
\end{tabular}
\end{table}

\section{Further Discussion} \label{sec:furtherDiscussion}
In this section, we briefly answer three common questions readers may have on the information-driven network calculus. First, why do we use our own definition on information and redundant information instead of using entropy and mutual information~\cite{Cov,sha} directly? Second, why do we use stochastic service curves and stochastic arrival curves instead of the deterministic ones~\cite{Le}? Third, there are other ways to define stochastic curves in the literature~\cite{Jia,Li}, why do we use the particular ones as in this paper? 

Regarding the first question, we emphasize that our definition does not conflict with Shannon's entropy and entropy rate~\cite{Cov,sha}. In a calculus for performance modeling, all we care is the properties of the information in a flow instead of its exact calculation. Due to this reason, we intentionally use generic notations, $H$ and $I$, to denote information and redundant information, which meet certain (intuitive) criteria, but leave their practical meaning and calculation open to users. This is to avoid different details in entropy estimation in particular applications~\cite{Lak,Lal}.  

Regarding the second question, it is commonly known that deterministic network calculus~\cite{Le} focuses on the \textit{worst case} analysis and as such the performance bounds obtained with deterministic network calculus is usually very loose. In addition, in many applications, there is uncertainty in both information processing and information transmission due to limited computational capacity (e.g., sensor nodes) and unreliable transmission links (e.g., wireless networks). For the information arrival process, although the entropy of a stationary process grows linearly with $t$ at a rate called entropy rate~\cite{Cov} and thus its information arrival curve could be modeled with a deterministic arrival curve, practically we usually do not know the entropy rate in advance and have to resort to \textit{sample entropy}~\cite{Lak,Lal}, which exhibits stochastic features.     

Lastly, it has been observed that there are other forms of definitions on stochastic service curves and stochastic arrival curves~\cite{Jia,Li}, for instance, the definitions with the \textit{sup} removed or with single \textit{sups}~\cite{Jia}. There are discussions in~\cite{Jia,Li} on the difficulties of deriving the basic properties of the calculus if other forms of definitions are used. Nevertheless, this does not necessarily mean that other definitions are improper. Actually we can transform one type of definition to another, but the constraint on the bounding function usually needs to change and the corresponding results have more complex form. Definitions \ref{def-sac} and \ref{def-ssc} have intentionally been chosen to ease the exposition.

%% Related Work
\section{Related Work}\label{sec:work}

To the best of our knowledge, there are currently no analytical tools available for systematic performance study of information driven-networks. The framework proposed in this paper is related to {\em network calculus}, particularly its stochastic branch: {\em stochastic network calculus}. Since its introduction in early 1990s~\cite{Cruz91a}, network calculus has attracted a lot of research attention and evolved along two tracks -- deterministic and stochastic. Excellent books summarizing results for deterministic network calculus are available (e.g.~\cite{Chang00,Le}). For stochastic network calculus, its research can also be tracked back to early 1990s (e.g.~\cite{Kurose92,Sidi93}). However, due to some difficulties specific to stochastic networks~\cite{Jia,Li}, it is only in recent years when critical network calculus properties such as concatenation property~\cite{Ciucu06,Jia} and independent case analysis~\cite{Jia} have been proved for stochastic network calculus. 

The relevance of the present paper to stochastic network calculus lies in the analogy between the various  models and properties defined or derived in this paper for information-driven networks, and the corresponding models and properties under stochastic network calculus. However, we want to stress that there is significant difference between the information network calculus proposed in the paper and the current development of stochastic network calculus, in that their targeted networks are different. The former is for information-driven networks where the key concern is about the quality of information delivery; the latter and network calculus in general are for networks where traffic is the focus. Due to this fundamental difference, special care has to be taken in developing the calculus for information-driven networks. For example, information dependence and information redundancy are unique concepts for information-driven networks. While two traffic flows may be independent, they can carry the same or highly correlated information as discussed in Example~\ref{ex:motivation}. \nop{The alert reader may have noticed that the current network calculus could be treated as a special case of the information network calculus developed in this paper. }

In information-driven networks, in-network information processing is likely performed. From this in-network processing viewpoint, the present paper is related to~\cite{Fidler06} and~\cite{Sch}. In~\cite{Fidler06},  scaling functions are used to model the relationship between the input traffic and the output traffic of a network element that processes the traffic. Based on the proposed scaling server model, \cite{Fidler06} extends the deterministic network calculus by considering data scaling in networks with in-network data processing. In~\cite{Sch}, how the scaling elements can be shifted across multiplexers in sensor networks is studied, which enables worst-case analysis of traffic delay and backlog in such networks. Note that in~\cite{Fidler06,Sch}, information processing only applies to intra-flow data, leaving inter-flow processing un-considered. However, in our work, both intra-flow processing and inter-flow processing are considered. In addition, the essential focus of~\cite{Fidler06} and~\cite{Sch} is on traffic, while our focus is on information carried by traffic. 

In~\cite{Ahl00}, the problem of {\em network information flow} is introduced. The focus of~\cite{Ahl00} is on a special type of in-network processing which is called {\em network coding}. With network coding, information is diffused through the network from the sources to the destination(s) and sources of flows may be jointly coded to achieve optimality in addressing the network information flow problem. An excellent introduction to network coding theory is available \cite{Yeung06}. \nop{In recent years, {\em network coding} has attracted a lot of research interest, and an update of the literature can be found from \cite{netcodehome}.} The present paper is related to~\cite{Ahl00} and network coding literature in that they all take {\em information} as the central point of study. 

Note that with network coding, sources may be coded jointly. In such cases, focusing on traffic in the analysis is no more applicable, since the output flow is not a simple scaling of the input flow or the aggregate of input flows. For example, suppose flows $f_1$ and $f_2$ are two bit streams. Applying exclusive-OR to the corresponding bits in them results in a {\em new} flow $f_1 \oplus f_2$. In the current network calculus literature including~\cite{Fidler06} and~\cite{Sch}, traffic amount and (traffic) service amount are the concern. In the example, for traffic, $A_{f_1 \oplus f_2}(t) = A_{f_1}(t) = A_{f_2}(t)$. For service to each individual flow or superposition of these two flows, however, the current network calculus approach provides no answer, since no corresponding output of $f_1$ or $f_2$ is found out of the exclusive-OR operation. With the proposed calculus in this paper, the exclusive-OR operation may be modeled as an information server providing a deterministic information service curve $\beta(t) = \infty$ to either flow $f_i$, $i=1, 2$. Then, based on results in this paper, particularly Theorem \ref{theorem:output}, we can say that the output flow $f_1 \oplus f_2$ preserves the information of either flow $f_i$, $i=1, 2$. 

Finally, we would like to highlight that network coding and other in-network processing techniques significantly complicate network performance analysis both in terms of traffic service guarantees and in terms of information service guarantees. While a lot of network calculus results are available for potential use in analyzing \textit{traffic} service guarantees in such networks (e.g.,~\cite{Mah}), no previous work has been found for analyzing \textit{information} service guarantees in these networks. We believe the proposed calculus makes a critical step and sheds light on further development to address this challenge. 

\section{Conclusions and Future Work}\label{sec:conclusion}
When network nodes become information aware and are capable of information processing, the network should not be simply considered as a data transportation tool but rather an information processing system. QoS guarantee in this type of networks should be measured with respect to quality of information instead of just data throughput or bounded (end-to-end) packet delay. Although substantial research has been done in information processing for specific applications, a systematic analytical framework for performance modeling and evaluation of information-driven networks remains blank. \nop{ due to two reasons. First, information processing is application dependent and cannot be captured easily with a generic framework. Second, the lack of effective control mechanisms due to resource constraints (e.g., sensor networks) makes the behavior of network nodes inherently stochastic, but the theory of stochastic network calculus makes clear progress only recently~\cite{Jia,Li}. }To the best of our knowledge, this paper is the first attempt to fill the gap by developing an analytical approach for performance evaluation of information-driven networks. We proved the basic properties and service guarantee of information-driven network calculus, derived the stochastically achievable information delivery rates, and investigated the problem of information transmission scheduling. 

This paper focuses mainly on the development of a new analytical framework, within which many interesting problems demand further investigation. These problems, for example, include the scheduling problem with a generic network topology for information fusion, the various optimization problems in communication as well as computation resource allocation, and the performance bounds if information error and lossy models are introduced.

\section*{Acknowledgment}
This work is supported by a fellowship from the Centre for Quantifiable Quality of Service in Communication Systems at the Norwegian University of Science and Technology, and by the Discovery Grant of Natural Sciences and Engineering Research Council of Canada.

\bibliographystyle{abbrv}
\balance
%\bibliographystyle{plain}
%\small \baselineskip 9pt
\bibliography{reference}

\begin{thebibliography}{10}

\bibitem{Ahl00}
R.~Ahlswede, N.~Cai, S.~Li, and R.~Yeung.
\newblock Network information flow.
\newblock {\em IEEE Trans. Information Theory}, 46(4):1204--1216, 2000.

\bibitem{Ahm}
R.~Ahmed and R.~Boutaba.
\newblock Distributed pattern matching: A key to flexible and efficient p2p
  search.
\newblock {\em IEEE Journal on Selected Areas in Communications}, 25(1):73--83,
  January 2007.

\bibitem{Arm}
G.~Armitage.
\newblock {\em Quality of Service in IP Networks}.
\newblock Macmillian TP, 2000.

\bibitem{Chang00}
C.-S. Chang.
\newblock {\em Performance Guarantees in Communication Networks}.
\newblock Springer-Verlag, 2000.

\bibitem{Ciucu06}
F.~Ciucu, A.~Burchard, and J.~Liebeherr.
\newblock A network service curve approach for the stochastic analysis of
  networks.
\newblock {\em IEEE Trans. Information Theory}, 52(6):2300--2312, June 2006.

\bibitem{Cov}
T.~M. Cover and J.~A. Thomas.
\newblock {\em Elements of Information Theory, Second Edition}.
\newblock John Wiley \& Sons, 2006.

\bibitem{Cruz91a}
R.~L. Cruz.
\newblock A calculus for network delay, part {I}: network elements in
  isolation.
\newblock {\em IEEE Trans. Information Theory}, 37(1):114--131, Jan. 1991.

\bibitem{Fidler06}
M.~Fidler and J.~B. Schmitt.
\newblock On the way to a distributed systems calculus: An end-to-end network
  calculus with data scaling.
\newblock In {\em Proc. ACM SIGMETRICS/Performance 2006}, pages 287--298, 2006.

\bibitem{Gar}
M.~Garey and D.~Johnson.
\newblock {\em Computers and Intractability: A Guide to the Theory of
  NP-Completeness}.
\newblock W.H. Freeman and Company, 1979.

\bibitem{Int}
C.~Intanagonwiwat, R.~Govindan, and D.~Estrin.
\newblock Directed diffusion: A scalable and robust communication paradigm for
  sensor networks.
\newblock In {\em Proceedings of ACM MobiCOM '00}, pages 56--67, Boston, August
  2000.

\bibitem{Jia}
Y.~Jiang.
\newblock A basic stochastic network calculus.
\newblock In {\em Proceedings of ACM Sigcomm 06}, pages 123--134, Pisa, Italy,
  September 2006.

\bibitem{Kun}
D.~Kundur, Z.~Liu, M.~Merabti, and H.~Yu.
\newblock Advances in peer-to-peer content search.
\newblock In {\em Proceedings of IEEE International Conference on Multimedia
  and Expo (ICME), 2007}, pages 404--407, Beijing, China, July 2007.

\bibitem{Kurose92}
J.~Kurose.
\newblock On computing per-session performance bounds in high-speed multi-hop
  computer networks.
\newblock In {\em ACM SIGMETRICS'92}, 1992.

\bibitem{Lak}
A.~Lakhina, M.~Croella, and C.~Diot.
\newblock Mining anomalies using traffic feature distributions.
\newblock In {\em Proceedings of ACM SIGCOMM 05}, pages 217--228, Philadelphia,
  USA, August 2005.

\bibitem{Lal}
A.~Lall, V.~Sekar, M.~Ogihara, J.~Xu, and H.~Zhang.
\newblock Data streaming algorithms for estimating entropy of network traffic.
\newblock {\em ACM SIGMETRICS Performance Evaluation Review}, 34(1):145--156,
  June 2006.

\bibitem{Le}
J.~{Le Boudec} and P.~Thiran.
\newblock {\em Network Calculus: A Theory of Deterministic Queuing Systems for
  the Internet}.
\newblock Springer-Verlag, 2001.

\bibitem{Li}
C.~Li, A.~Burchard, and J.~Liebeherr.
\newblock A network calculus with effective bandwidth.
\newblock {\em IEEE/ACM Transactions on Networking}, 15(6):1142--1453, December
  2007.

\bibitem{Liu}
C.~Liu, K.~Wu, and J.~Pei.
\newblock An energy efficient data collection framework for wireless sensor
  networks by exploiting spatiotemporal correlation.
\newblock {\em IEEE Transactions on Parallel and Distributed Systems (TPDS)},
  18(7):1010--1023, July 2007.

\bibitem{Mah}
A.~Mahmino, J.~Lecan, and C.~Fraboul.
\newblock Guaranteed packet delays with network coding.
\newblock In {\em First IEEE International Workshop on Wireless Network
  Coding}, pages 1--6, San Francisco, June 2008.

\bibitem{Pat}
S.~Pattem, B.~Krishnamachari, and R.~Govindan.
\newblock The impact of spatial correlation on routing with compression in
  wireless sensor networks.
\newblock {\em ACM Transactions on Sensor Networks (TOSN)}, 4(4):article 24,
  August 2008.

\bibitem{Sch}
J.~B. Schmitt, F.~A. Zdarsky, and L.~Thiele.
\newblock A comprehensive worst-case calculus for wireless sensor networks with
  in-network processing.
\newblock In {\em Proceedings of IEEE RTSS 2007}, pages 193--202, Tucson,
  Arizona, December 2007.

\bibitem{sha}
C.~E. Shannon.
\newblock A mathematical theory of communication.
\newblock {\em Bell System Technical Journal}, 27:379--423,623--656, 1948.

\bibitem{Sze}
R.~Szewczyk, J.~Polastre, A.~Mainwaring, and D.~Culler.
\newblock Lessons from a sensor network expedition.
\newblock In {\em Proceedings of the First European Workshop on Sensor Networks
  (EWSN)}, pages 307--322, Berlin, Germany, January 2004.

\bibitem{Vur}
M.~Vuran, O.~B. Akan, and I.~Akyildiz.
\newblock Spatio-temporal correlation: Theory and applications for wireless
  sensor networks.
\newblock {\em Computer Networks}, 45(3):245--259, June 2004.

\bibitem{Sidi93}
O.~Yaron and M.~Sidi.
\newblock Performance and stability of communication network via robust
  exponential bounds.
\newblock {\em IEEE/ACM Trans. Networking}, 1(3):372--385, June 1993.

\bibitem{Yeung06}
R.~Yeung, S.~Li, N.~Cai, and Z.~Zhang.
\newblock {\em Network Coding Theory}.
\newblock now Publishers, 2006.

\end{thebibliography}

\newpage
\small \baselineskip 9pt
\appendix
%Appendix A
\section{Proofs of Results} 
\label{app:A}
\textbf{Proof of Lemma~\ref{lemma:2}.} For any random variables $X$ and $Y$, and any $x \ge 0, z \ge 0$, $\{X-Y \ge x\} \cap \{X \le x+z\} \cap \{Y > z\} = \emptyset$, where $\emptyset$ denotes the null set. We thus have 
$$\{X-Y \ge x\} \subseteq \{X > x+z\} \cup \{Y \le z\},$$ 
which means
$$Prob\{X-Y \ge x\} \le Prob\{X > x+z\}+ Prob\{Y \le z\}.$$ 
Since the above inequality holds for all $z \ge 0$, we get 
$$Prob\{X-Y \ge x\} \le \inf_{z\ge 0}[Prob\{X > x+z\}+ Prob\{Y \le z\}],$$ 
with which and $\bar{F}_X(x) \le f(x)$ and $F_Y(x) \le g(x)$, where $f\in \bar{\mathcal{F}}, g\in \mathcal{F}$, the result is proved.\QED \\

\textbf{Proof of Theorem~\ref{theorem:superposition} (Superposition).} Based on the properties of information and redundant information, we have 
\begin{align*}
&\sup_{0\le u \le s} [H(A(u, s)) - (\alpha_1(s-u)+\alpha_2(s-u)-\gamma(s-u))] \\
= &\sup_{0\le u \le s} [H(A_1(u,s)) + H(A_2(u,s)) - I(A_1;A_2)(u,s) - \\
& (\alpha_1(s-u)+\alpha_2(s-u)-\gamma(s-u))] \\ 
\le &\sup_{0\le u \le s}[H(A_1(u,s))-\alpha_1(s-u)] % \\&
+ \sup_{0\le u \le s}[H(A_2(u,s))-\alpha_2(s-u)] \\
& - \inf_{0\le u \le s}[I(A_1;A_2)(u,s)-\gamma(s-u)]
\end{align*}
for any $s\ge 0$, from which, we further get
\begin{align*}
&\sup_{0\le s \le t}\sup_{0\le u \le s} [H(A(u, s)) - (\alpha_1(s-u)+\alpha_2(s-u)-\gamma(s-u))] \\
\le &\sup_{0\le s \le t}\sup_{0\le u \le s}[H(A_1(u,s))-\alpha_1(s-u)] + \\
& \sup_{0\le s \le t}\sup_{0\le u \le s}[H(A_2(u,s))-\alpha_2(s-u)] \\
& - \inf_{0\le s \le t}\inf_{0\le u \le s}[I(A_1;A_2)(u,s)-\gamma(s-u)].
\end{align*}
%\begin{align*}
%&\sup_{0\le s \le t} [H(A(s,t)) - (\alpha_1(t-s)+\alpha_2(t-s)-\gamma(t-s))] \\
%= &\sup_{0\le s \le t} [H(A_1(s,t)) + H(A_2(s,t)) - I(A_1;A_2)(s,t) - \\
%& (\alpha_1(t-s)+\alpha_2(t-s)-\gamma(t-s))] \\ 
%\le &\sup_{0\le s \le t}[H(A_1(s,t))-\alpha_1(t-s)] + \\
%& \sup_{0\le s \le t}[H(A_2(s,t))-\alpha_2(t-s)] \\
%& - \inf_{0\le s \le t}[I(A_1;A_2)(s,t)-\gamma(t-s)]
%\end{align*}

From the above inequality, the theorem is proved with the definitions of information stochastic arrival curve and low-bounded information stochastic arrival curve, Lemma~\ref{lemma:1}, and Lemma~\ref{lemma:2}. \QED\\

\textbf{Proof of Theorem~\ref{theorem:concatenation} (Concatenation).} We only prove the two node case, because the same result can be extended to the $N$-node case. We use $A^i(t)$ and $A^{i*}(t)$ to denote the input flow and the out flow of node $i$, respectively. Note that $A^{1*}(t) = A^2(t).$ For any $s\ge 0$ and $x \ge 0$, we have 
\begin{align*}
&H(A^1(s))\otimes[\beta^1\otimes \beta^2]^{x}(s)-H(A^{2*}(s))\\
\le &H(A^1(s))\otimes ([\beta^1]^{x}\otimes [\beta^2]^{x})(s)-H(A^{2*}(s))\\
= &\inf_{0\le u \le s}[H(A^1(u))\otimes[\beta^1]^{x}(u)+[\beta^2]^{x}(s-u)-H(A^{1*}(u)) \\
 & + H(A^2(u))] - H(A^{2*}(s))\\
\le & \sup_{0\le u\le s}[H(A^1(u))\otimes[\beta^1]^{x}(u) - H(A^{1*}(u))] \\
   & + \inf_{0\le u \le s}[H(A^2(u))+ [\beta^2]^{x}(s-u)] - H(A^{2*}(s)) \\ 
= & \sup_{0\le u\le s}[H(A^1(u))\otimes[\beta^1]^{x}(u)-H(A^{1*}(u))] \\
   & + H(A^2(s))\otimes[\beta^2]^{x}(s) - H(A^{2*}(s)). \\ 
\end{align*} 
%\begin{align*}
%&H(A^1(s))\otimes(\beta^1\otimes \beta^2)(s)-H(A^{2*}(s))\\
%= &\inf_{0\le u \le s}[H(A^1(u))\otimes\beta^1(u)+\beta^2(s-u)-H(A^{1*}(u)) \\
% & + H(A^2(u))] - H(A^{2*}(s))\\
%\le & \sup_{0\le u\le s}[H(A^1(u))\otimes\beta^1(u) - H(A^{1*}(u))] \\
%   & + \inf_{0\le u \le s}[H(A^2(u))+ \beta^2(s-u)] - H(A^{2*}(s)) \\ 
%= & \sup_{0\le u\le s}[H(A^1(u))\otimes\beta^1(u)-H(A^{1*}(u))] \\
%   & + H(A^2(s))\otimes\beta^2(s) - H(A^{2*}(s)). \\ 
%\end{align*} 
We thus have for any $t\ge 0$, 
\begin{align*}
&\sup_{0\le s \le t}[H(A^1(s))\otimes[\beta^1\otimes \beta^2]^{x}(s)-H(A^{2*}(s))]\\
\le & \sup_{0\le u\le t}[H(A^1(u))\otimes[\beta^1]^{x}(u)-H(A^{1*}(u))] \\
   & + \sup_{0\le s \le t}[H(A^2(s))\otimes[\beta^2]^{x}(s) - H(A^*(s))]. 
\end{align*}
The theorem is proved with the above inequality, the definition of information stochastic service curve, and Lemma~\ref{lemma:1}. \QED\\

\textbf{Proof of Theorem~\ref{theorem:output} (Output).} Based on the properties of information, for any $0\le s \le t$, we have
\begin{align*}
&H(A^*(s,t)) = H(A^*(t))-H(A^*(s)) \\
\le & H(A(t)) - H(A^*(s)) \\
= &H(A(t)) - H(A(s))\otimes [\beta]^{x}(s) + H(A(s))\otimes [\beta]^{x}(s) %\\&
  - H(A^*(s))\\
= &\sup_{0\le u \le s}[H(A(t)) - H(A(u))- [\beta]^{x}(s-u)] \\
& + [H(A(s))\otimes [\beta]^{x}(s) - H(A^*(s))]\\ 
 = &\sup_{0\le u \le s}[H(A(u,t)) - \alpha(t-u) + \alpha (t-u) - [\beta]^{x}(s-u)]\\
 & + [H(A(s))\otimes [\beta]^{x}(s) - H(A^*(s))]\\
 \le & \sup_{0\le u \le t}[H(A(u,t)) - \alpha(t-u)] + \alpha\oslash\beta(t-s) \\
 & + [H(A(s))\otimes [\beta]^{x}(s) - H(A^*(s))] 
\end{align*}
We thus have 
\begin{align*}
&\sup_{0\le s \le t}[H(A^*(s,t)) - \alpha\oslash\beta(t-s)]\\
\le & \sup_{0\le u \le t}[H(A(u,t))-\alpha(t-u)] %\\&
    + \sup_{0\le s \le t}[H(A(s))\otimes [\beta]^{x}(s) - H(A^*(s))] 
\end{align*}
from which together with simple manipulation, we further get
\begin{align*}
&\sup_{0\le s \le t}\sup_{0\le u \le s}[H(A^*(u,s)) - \alpha\oslash\beta(s-u)]\\
\le & \sup_{0\le s \le t}\sup_{0\le u \le s}[H(A(u,s))-\alpha(s-u)]\\
    & + \sup_{0\le s \le t}[H(A(s))\otimes [\beta]^{x}(s) - H(A^*(s))].
\end{align*}

From the above inequality, the theorem is proved with the definition of information stochastic arrival curve, the definition of information stochastic service curve, and Lemma~\ref{lemma:1}. \QED\\

\textbf{Proof of Theorem~\ref{theorem:service} (Service guarantee).} 1) For the information backlog $B(t)$, we have for any $t, x\ge 0$, 
\begin{align*}
&B(t)= H(A(t))-H(A^*(t)) \\ 
=& H(A(t))- H(A(t))\otimes[\beta]^{x}(t) + H(A(t))\otimes[\beta]^{x}(t)- H(A^*(t)) \\
=&\sup_{0\le s\le t}[H(A(s,t))-\alpha(t-s) + \alpha(t-s) -[\beta]^{x}(t-s)] \\
 &  +[H(A(t))\otimes[\beta]^{x}(t)- H(A^*(t))]\\
\le &\sup_{0\le s\le t}[H(A(s,t))-\alpha(t-s)] + \alpha\oslash\beta(0)\\
&+[H(A(t))\otimes[\beta]^{x}(t)- H(A^*(t))] \\
\le &\sup_{0\le s\le t}[H(A(s,t))-\alpha(t-s)] + \alpha\oslash\beta(0)\\
&+\sup_{0\le s\le t}[H(A(s))\otimes[\beta]^{x}(s)- H(A^*(s))] 
\end{align*}
The result is proved with the above inequality, the definition of information stochastic arrival curve, the definition of information stochastic service curve, and Lemma~\ref{lemma:1}.

2) For the information delay $D(t)$, we have from the definition, for any $y \ge 0$, $\{D(t) > y\} \subset \{H(A(t))>H(A^*(t+y))\}$ and hence $Prob\{D(t) > y\} \le Prob\{H(A(t))>H(A^*(t+y))\}$. We also have:  
\begin{align*}
&H(A(t))-H(A^*(t+y)) \\ 
=& H(A(t))- H(A(t+y))\otimes[\beta]^{x}(t+y) + H(A(t+y))\otimes[\beta]^{x}(t+y)\\
  & +\alpha(t-s) - \alpha(t-s) - H(A^*(t+y)) \\
\le &\sup_{0\le s\le t}[H(A(s,t))-\alpha(t-s)] \\
    & + H(A(t+y))\otimes[\beta]^{x}(t+y) - H(A^*(t+y)) \\
    & + \sup_{0 \le s\le t+y}[\alpha(t-s) -[\beta]^{x}(t-s+y)]
\end{align*}
By replacing $y$ with $h(\alpha+{x},[\beta]^{x})$ in above, where $h(\alpha+{x},[\beta]^{x})=\sup_{s>0}\{\inf\{\tau\ge 0: \alpha(s)+{x} \le [\beta]^{x}(s+\tau)\}\}$ is the maximum horizontal distance between functions $\alpha(t)+{x}$ and $[\beta]^{x}(t)$, which implies $\alpha(t)+x \le [\beta]^{x}(t + h(\alpha +{x}, [\beta]^{x}))$, we obtain
%If we define $\beta'(t) = \max\{\beta(t), x\}$, by replacing $x$ with $h(\alpha+y,\beta)$ where $h(\alpha,\beta)=\sup_{s>0}\{\inf\{\tau\ge 0: \alpha(s)\le\beta(s+\tau)\}\}$ is the maximum horizontal distance between functions $\alpha$ and $\beta$, we obtain
\begin{align*}
&H(A(t))-H(A^*(t+h(\alpha+x,[\beta]^{x}))) \\ 
\le & \sup_{0\le s\le t}[H(A(s,t))-\alpha(t-s)] \\
    & + H(A(t+h(\alpha+x,[\beta]^{x})))\otimes[\beta]^{x}(t+h(\alpha+x,[\beta]^{x})) \\
    & - H(A^*(t+h(\alpha+x,[\beta]^{x})))-x \\
\le & \sup_{0\le s\le t}[H(A(s,t))-\alpha(t-s)] + \\
    & \sup_{0\le s\le t}[H(A(s+h(\alpha+x,[\beta]^{x})))\otimes[\beta]^{x}(s+h(\alpha+x,[\beta]^{x})) \\
    & - H(A^*(s+h(\alpha+x,[\beta]^{x})))]-x     
\end{align*}
Based on the above inequality, the definition of information stochastic arrival curve, the definition of information stochastic service curve, and Lemma~\ref{lemma:1}, we have $Prob\{D(t)\}> h(\alpha+x,[\beta]^{x})\} \le f\otimes g(x).$ 
%Based on the above inequality, the definition of information stochastic arrival curve, the definition of information stochastic service curve, and Lemma~\ref{lemma:1}, we have $Prob\{D(t)\}> h(\alpha+x,\beta')\} \le f\otimes g(x).$ 

3) For the information backlog within delay bound $\tau (\le D(t))$, using the same derivation as in (2), we have 
\begin{align*}
&\hat{B}(t,\tau)= H(A(t))-H(A^*(t+\tau)) \\ 
\le &\sup_{0\le s\le t}[H(A(s,t))-\alpha(t-s)] \\
    & + H(A(t+\tau))\otimes[\beta]^{x}(t+\tau) - H(A^*(t+\tau)) \\
    & + \sup_{0 \le s\le t+\tau}[\alpha(t-s) -[\beta]^{x}(t-s+\tau)] \\
 = & \sup_{0\le s\le t}[H(A(s,t))-\alpha(t-s)] \\
    & + H(A(t+\tau))\otimes[\beta]^{x}(t+\tau) - H(A^*(t+\tau)) \\
    & - \inf_{v\ge 0}[[\beta]^{x}(v)-\alpha(v-\tau)]  \\
 \le & \sup_{0\le s\le t}[H(A(s,t))-\alpha(t-s)] \\
    & + \sup_{0\le s\le t}[H(A(s+\tau))\otimes[\beta]^{x}(s+\tau) - H(A^*(s+\tau))] \\
    & - \inf_{v\ge 0}[\beta(v)-\alpha(v-\tau)]       
\end{align*}
The result is proved with the above inequality, the definition of information stochastic arrival curve, the definition of information stochastic service curve, and Lemma~\ref{lemma:1}.\QED\\

\textbf{Proof of Theorem~\ref{theorem:impair} (Service reduction with impairment).} We could treat the system as if it provides service to the ``aggregate" of two flows: the input $A(t)$ and the impairment $\hat{\mathcal{I}}(t)$. Denote the ``aggregate" as $\hat{A}(t)$. There holds $H(\hat{A}(t)) = H(A(t)) + H(\hat{\mathcal{I}}(t))$ and $H(\hat{A}^*(t)) = H(A^*(t)) + H(\hat{\mathcal{I}^*}(t))$, where $\hat{A}^*(t), A^*(t)$, and $\hat{\mathcal{I}^*}(t)$ are the outputs of $\hat{A}(t), A(t)$, and $\hat{\mathcal{I}}(t)$, respectively. We have for any $s, x \ge 0$,
\begin{align*}
& H(A(s)) \otimes [\beta-\alpha]^{x} (s) - H(A^*(s)) \\
& \le \inf_{0\le u \le s}[H(\hat{A}(u)) - H(\hat{\mathcal{I}}(u))+ [\beta]^{x}(s-u)-\alpha(s-u)] \\
& - H(\hat{A}^*(s)) + H(\hat{\mathcal{I}^*}(s)) \\
& \le [H(\hat{A}(s))\otimes [\beta]^{x}(s) - H(\hat{A}^*(s))]  - \inf_{0\le u \le s}[H(\hat{\mathcal{I}}(u))+\alpha(s-u)] \\
&+ H(\hat{\mathcal{I}^*}(s)) \\
& \le [H(\hat{A}(s))\otimes [\beta]^{x}(s) - H(\hat{A}^*(s))]  - \inf_{0\le u \le s}[H(\hat{\mathcal{I}}(u))+\alpha(s-u)] \\
& + H(\hat{\mathcal{I}}(s)) \\
& =  [H(\hat{A}(s))\otimes [\beta]^{x}(s) -H(\hat{A}^*(s))] + \sup_{0\le u \le s} [H(\hat{\mathcal{I}}(u,s)) - \alpha(s-u)]   
\end{align*} 
We hence have for any $t\le 0$,
\begin{align*}
& \sup_{0\le s \le t}[H(A(s)) \otimes [\beta-\alpha]^{x} (s) - H(A^*(s))] \\
& \le \sup_{0\le s \le t}[H(\hat{A}(s))\otimes [\beta]^{x}(s) -H(\hat{A}^*(s))] \\&
+ \sup_{0\le s \le t}\sup_{0\le u \le s} [H(\hat{\mathcal{I}}(u,s)) - \alpha(s-u)].   
\end{align*} 
The theorem is proved because the system provides $\hat{A(t)}$ with service $\sim_{i.s.s.}<g,\beta>$ and the impairment process $\hat{\mathcal{I}}$ follows $\sim_{i.s.a.}<f,\alpha>$.\QED \\

\textbf{Proof of Theorem~\ref{theorem:parallel} (Parallel servers).}  We only prove the case of two parallel nodes, because the result can be easily extended to the case of $N$ parallel nodes. Because sub-flows are information exclusive, we have $H(A(t))=H(A_1(t))+H(A_2(t))$. Let $\beta(t) = \beta^1(t)+\beta^2(t)$. Because the weight assigned to sub-flow $A_i$ in the weighted information splitter is $\beta^i$, for any $s > 0$, $\frac{H(A_1(s))}{H(A(s))} = \frac{\beta^1(s)}{\beta(s)} \equiv \lambda_1(s)$, $\frac{H(A_2(s))}{H(A(s))} = \frac{\beta^2(s)}{\beta(s)} \equiv \lambda_2(s)$, where $\lambda_1(s)+\lambda_2(s)=1$.  Therefore, for any $s\ge 0$, we have: (Due to space limitation, in the following derivation, we simply use $\beta$, $\beta^1$, $\beta^2$ to respectivaly represent $[\beta]^{x}$, $[\beta^1]^{x}$ and $[\beta^2]^{x}$.)
\begin{eqnarray}
& &H(A(s))\otimes \beta(s) - H(A^*(s)) \nonumber \\ 
&=& H(A(s))\otimes \beta(s) - H(A_1(s))\otimes \beta^1(s)- H(A_2(s))\otimes \beta^2(s) \nonumber \\
& & + [H(A_1(s))\otimes \beta^1(s) -H(A_1^*(s))] \nonumber \\
 & & + [H(A_2(s))\otimes \beta^2(s) -H(A_2^*(s))]. \label{eq:para}
\end{eqnarray}
Looking at the first three items, we have 
\begin{eqnarray}
& & H(A(s))\otimes \beta(s) - H(A_1(s))\otimes \beta^1(s)- H(A_2(s))\otimes \beta^2(s) \nonumber \\
&=&H(A(s))\otimes \beta(s) -(\lambda_1(s)H(A(s)))\otimes (\lambda_1(s)\beta(s)) \nonumber \\
&  & - (\lambda_2(s)H(A(s)))\otimes (\lambda_2(s)\beta(s)) \nonumber \\
&=&\inf_{0\le u \le s} [H(A(u)) + \beta(s-u)] \nonumber \\
& & - \inf_{0\le u \le s} [\lambda_1(u)H(A(u)) + \lambda_1(s-u)\beta(s-u)]\nonumber \\
& & - \inf_{0\le u \le s} [\lambda_2(u)H(A(u)) + \lambda_2(s-u)\beta(s-u)] \label{eq:para2}
\end{eqnarray}
Assume that $\inf_{0\le u \le s} [\lambda_1(u)H(A(u)) + \lambda_1(s-u)\beta(s-u)] = \lambda_1(u_1)H(A(u_1)) + \lambda_1(s-u_1)\beta(s-u_1)$, $\inf_{0\le u \le s} [\lambda_2(u)H(A(u)) + \lambda_2(s-u)\beta(s-u)] = \lambda_2(u_2)H(A(u_2)) + \lambda_2(s-u_2)\beta(s-u_2)$. Without loss of generality, assume that $0\le u_1\le u_2\le s$. 

%\begin{itemize}
%\item Case 1: If during the period $(u_1, u_2]$, $H(A(u_1,u_2)) \ge \beta(u_1-u_2)$, Equation~(\ref{eq:para2}) is
Case 1: If during the period $(u_1, u_2]$, $H(A(u_1,u_2)) \ge \beta(u_1-u_2)$, Equation~(\ref{eq:para2}) is
\begin{align*}
\le & H(A(u_1)) + \beta(s-u_1) - \lambda_1(u_1)H(A(u_1)) - \lambda_1(s-u_1)\beta(s-u_1) \\
   &  -\lambda_2(u_2)H(A(u_2)) - \lambda_2(s-u_2)\beta(s-u_2)\\
=& \lambda_2(u_2-u_1)[\beta(u_2-u_1) - H(A(u_1,u_2))] \le 0    
\end{align*}

%\item Case 2: If during the period $(u_1, u_2]$, $H(A(u_1,u_2)) \le \beta(u_1-u_2)$, Equation~(\ref{eq:para2}) is
Case 2: If during the period $(u_1, u_2]$, $H(A(u_1,u_2)) \le \beta(u_1-u_2)$, Equation~(\ref{eq:para2}) is
\begin{align*}
\le & H(A(u_2)) + \beta(s-u_2) - \lambda_1(u_1)H(A(u_1)) - \lambda_1(s-u_1)\beta(s-u_1) \\
   &  -\lambda_2(u_2)H(A(u_2)) - \lambda_2(s-u_2)\beta(s-u_2)\\
=& \lambda_1(u_2-u_1)[H(A(u_1,u_2))-\beta(u_2-u_1)] \le 0    
\end{align*}
%\end{itemize}

Therefore, Equation~(\ref{eq:para2}) is no larger than $0$ in any case. For $\forall t \ge 0$, Equation~(\ref{eq:para}) thus becomes:
\begin{align*}
&\sup_{0\le s \le t}[H(A(s))\otimes (\beta^1+\beta^2)(s) - H(A^*(s))]  \\
\le & \sup_{0\le s \le t}[H(A_1(s))\otimes \beta^1(s) -H(A_1^*(s))]\\
   & + \sup_{0\le s \le t}[H(A_2(s))\otimes \beta^2(s) -H(A_2^*(s))]
\end{align*}
The theorem is hence proved with the above inequality, the definition of information stochastic service curve, and Lemma~\ref{lemma:1}. \QED\\

\textbf{Proof of Theorem~\ref{theorem:correlation}.} We have     
\begin{align*}
H(A(t))& = H(\Delta_1(t)) + H(\Delta_2(t)) %\\&
        - I(\Delta_1(t); \Delta_2(t)),
\end{align*}
and 
\begin{align*}
H(A^*(t+\tau))& = H(\Delta^*_1(t+\tau)) + H(\Delta^*_2(t+\tau)) \\
       & - I(\Delta^*_1(t+\tau); \Delta^*_2(t+\tau)).
\end{align*}
The theorem is obvious because  $H(A(t))= H(A^*(t+\tau))$, $H(\Delta^*_1(t+\tau)) = H(\Delta_1(t))$, and $H(\Delta^*_2(t+\tau)) = H(\Delta_2(t)).$\QED

\end{document}